\documentclass[twocolumn,secnumarabic,amssymb, nobibnotes, aps, pra,reprint]{revtex4-2}

\usepackage{braket}
\usepackage{amsmath}
\usepackage{graphicx}
\usepackage{xcolor}
\usepackage[scr=boondox]{mathalpha} 

\usepackage{hyperref}
\hypersetup{
    colorlinks,
    citecolor=blue,
    filecolor=red,
    linkcolor=magenta,
    urlcolor=red
}

\newcommand{\tr}[1]{\text{Tr}\left[ #1\right]} 
\newcommand{\bo}[1]{\hat{\boldsymbol #1}} 
\renewcommand{\braket}[1]{\left\langle #1 \right\rangle} 
\newcommand{\bs}[1]{\boldsymbol{#1}} 

\makeatletter
\newcommand*{\overrightharpoonup}{\mathpalette{\overarrow@\varrightharpoonupfill@}}
\newcommand*{\varrightharpoonupfill@}{\arrowfill@\relbar\relbar\varrightharpoonup}
\makeatother

\DeclareFontFamily{U}{matha}{\hyphenchar\font45}
\DeclareFontShape{U}{matha}{m}{n}{
      <5> <6> <7> <8> <9> <10> gen * matha
      <10.95> matha10 <12> <14.4> <17.28> <20.74> <24.88> matha12
      }{}
\DeclareSymbolFont{matha}{U}{matha}{m}{n}
\DeclareMathSymbol{\varrightharpoonup}{3}{matha}{"E1}

\begin{document}

\title{Laser cooling trapped-ion crystal modes beyond the Lamb-Dicke regime}

\author{John P. Bartolotta}
\email{john.bartolotta@quantinuum.com}
\affiliation{Quantinuum, 303 S. Technology Ct., Broomfield, CO 80021, USA}
\author{Brian Estey}
\affiliation{Quantinuum, 303 S. Technology Ct., Broomfield, CO 80021, USA}
\author{Michael Foss-Feig}
\affiliation{Quantinuum, 303 S. Technology Ct., Broomfield, CO 80021, USA}
\author{David Hayes}
\affiliation{Quantinuum, 303 S. Technology Ct., Broomfield, CO 80021, USA}
\author{Christopher N. Gilbreth}
\email{christopher.gilbreth@quantinuum.com}
\affiliation{Quantinuum, 303 S. Technology Ct., Broomfield, CO 80021, USA}
\date{\today}

\begin{abstract}
Laser cooling methods for trapped ions are most commonly studied at low energies, i.e., in the Lamb-Dicke regime.
However, ions in experiments are often excited to higher energies for which the Lamb-Dicke approximation breaks down.
Here we construct a non-perturbative, semiclassical method for predicting the energy-dependent cooling dynamics of trapped-ion crystals with potentially many internal levels and motional modes beyond the Lamb-Dicke regime.
This method allows accurate and efficient modeling of a variety of interesting phenomena, such as the breakdown of EIT cooling at high energies and the simultaneous cooling of multiple high-temperature modes.
We compare its predictions both to fully-quantum simulations and to experimental data for a broadband EIT cooling method on a Raman $S$-$D$ transition in $^{138}$Ba${^+}$. 
We find the method can accurately predict cooling rates over a wide range of energies relevant to trapped ion experiments. 
Our method complements fully quantum models by allowing for fast and accurate predictions of laser-cooling dynamics at much higher energy scales.
\end{abstract}

\maketitle

\section{Introduction}

Certain applications of trapped ions, such as quantum information processing~\cite{cirac1995} and optical clocks~\cite{brewer2019}, require the removal of excess thermal energy in the ionic motion.
These applications 
often present the challenge of designing laser-cooling protocols~\cite{eschner2003} which are effective over potentially many orders of magnitude in energy and in the presence of heat-inducing processes, such as ion transport~\cite{delaney2024} and collisions with background gas particles~\cite{hankin2019,Mourik2021}.
It is therefore desirable for a theoretical model to accurately and efficiently predict cooling dynamics over a wide energy range.

In some treatments~\cite{Stenholm1986}, the cooling dynamics are perturbatively expanded in the Lamb-Dicke parameter $\eta$, which requires excursions of the ion's position to be small compared to the wavelength $\lambda$ of the cooling light (the so-called Lamb-Dicke regime). 
Another approximation is to assume a separation of timescales between the internal and motional dynamics such that the internal dynamics can be adiabatically eliminated, resulting in equations of motion on the motional subspace in the form of Newton's equations~\cite{metcalfText}, time-averaged energy rate equations \cite{fluorescenceTheory}, Fokker-Planck equations~\cite{dalibard1985,molmer1994,morigi1999,morigi2001,rabl2010,highTempOscillator}, or population rate equations~\cite{Cirac1992,morigi2000,Rasmusson2021}. 
However, this separation of timescales is not always satisfied and may not be the ideal operating regime for efficient cooling.
Alternatively, the quantum properties of the motion can be neglected in favor of a classical treatment~\cite{Stenholm1986,rabl2010}, at the expense of losing certain (typically negligible) quantum correlations~\cite{dalibard1985}, to derive a ``semiclassical" model that evolves a set of phase space coordinates $\{\bs r, \bs p\}$.
In contrast to the Lamb-Dicke regime treatment, the semiclassical approach only requires that the width of the ionic wavepacket is small compared to $\lambda$. As such, it can provide complementary information at high temperatures where the Lamb-Dicke treatment is not valid.

In this work, we present and test a generalized semiclassical method to calculate the average rate of energy change in a trapped-ion crystal interacting with one or more laser fields.
The internal dynamics are evolved through a quantum master equation, and the internal and motional dynamics are coupled through the laser phases witnessed by the ion~\cite{cook1979,fluorescenceTheory}.
Although this approach generally precludes an analytic solution, we show that it provides a computationally efficient model of laser cooling that is valid over a broad parameter range.
The method is simple to implement for multi-ion systems, each with potentially many relevant internal levels, and because it makes no expansion in $\eta$, it is accurate beyond the Lamb-Dicke regime. 
Further, it makes no assumptions about the relative size of trap periods $T$ and internal state lifetimes $\tau$, and so is applicable beyond the ``weak-binding" regime $\tau \ll T$~\cite{fluorescenceTheory,metcalfText} at the expense of explicitly tracking the internal dynamics.
We apply this method to predict cooling rates across a broad range of energies for an electromagnetically-induced transparency (EIT) cooling protocol~\cite{morigi2000} we propose for broadband sympathetic cooling using  $^{138}$Ba$^+$, and compare its predictions both to a fully-quantum master equation and to experiment.
In addition to exploring a practical parameter regime, this comparison highlights EIT cooling's rich structure as a function of mode energy, which poses a significant challenge to the unification of laser-cooling theory over a broad energy range and therefore serves as a stringent benchmark of the semiclassical model.

We begin by presenting the derivation of a semiclassical cooling rate in Section~\ref{semiRates}.
In Section~\ref{quantumComparison}, we demonstrate the accuracy of the semiclassical model in the context of EIT cooling in a $\Lambda$-system by comparing its predictions to well-studied quantum models.
In Section~\ref{expComparison}, we compare the predictions of the model to a cooling experiment.
In Section~\ref{instantaneousRates}, we demonstrate how to use the semiclassical cooling rates to calculate dynamics for a more general class of motional distributions.
In Section~\ref{multiMode}, we consider a scenario when multiple modes are cooled simultaneously. 

\section{Calculation of Semiclassical Cooling Rates}
\label{semiRates}

Here we outline the essential features of our semiclassical model for which we provide a detailed derivation in Appendix~\ref{modelDerivation}.
Then we derive the semiclassical cooling rate which we use in the following sections.

The key ingredient of the semiclassical approximation is to replace all instances of the quantum motional operators with their expectation values $\bs r_j(t) = \braket{\hat{\bs r}_j(t)}$ and $\bs p_j(t) = \braket{\hat{\bs p}_j(t)}$ for each ion $j$ and evolve these quantities according to classical equations of motion including dissipative forces from the lasers.
For simplicity, we perform the rotating wave approximation for the laser-ion interactions and ignore any inter-ion internal state quantum coherences, absorption of spontaneously emitted photons by neighboring ions, and recoil from spontaneously emitted photons.
With these approximations, we arrive at the classical equations of motion
\begin{align}
\label{Newton}
\begin{split}
	\frac{d \bs{r}_j}{dt} & = \frac{\bs{p}_j}{m_j}; \\
	\frac{d \bs{p}_j}{dt} & = \bs F_j^\text{laser} + \bs F_j^\text{trap} + \bs F_j^\text{Coloumb}
\end{split}
\end{align}
and quantum master equations ($\hbar \equiv 1$)
\begin{equation}
\label{internalME}
	\frac{d \hat \rho^I_j}{dt} =  -i \left[ \hat H_j(\bs r_j, t) , \, \hat \rho^I_j \right]
	+ \mathcal{D}_j\hat \rho^I_j.
\end{equation}
In Eq.~\eqref{internalME}, $\hat \rho^I_j$ is the reduced density matrix on the internal subspace of the $j^\text{th}$ ion, $\mathcal{D}_j\hat \rho^I_j$ accounts for dissipative dynamics on the internal subspace, and
\begin{equation}
	\hat H_j(\bs r_j, t) = \sum_{l,\alpha,\beta} \frac{\Omega_{\alpha \beta}^{(l,j)}}{2} \hat{\sigma}_{\alpha \beta}^{(j)} e^{- i \phi^{(l,j)}_{\alpha \beta}(\bs r_j,t)} + \text{h.c.}
\end{equation}
is the ion-laser interaction Hamiltonian.
In the Hamiltonian, $\Omega_{\alpha \beta}^{(l,j)}$ is the Rabi frequency between internal states $\ket{\alpha},\ket{\beta}$ of ion $j$ induced by laser $l$, $\hat{\sigma}_{\alpha \beta}^{(j)}$ is a transition operator, and
\begin{equation}
\phi^{(l,j)}_{\alpha \beta}(\bs r_j, t) =  \int_{0}^t  \left[\Delta_{\alpha \beta}^{(l)}(t') - \bs k_l \cdot \bs v_j(t')\right]  dt'
\end{equation}
is the instantaneous phase of laser $l$ witnessed by ion $j$.
This phase depends on the laser detuning $\Delta_{\alpha \beta}^{(l)}(t)$ and (crucially) the Doppler shift $\bs k_l \cdot \bs v_j(t)$ for an ion with velocity  $\bs v_j(t)$ interacting with a laser with wavevector $\bs k_l$, thereby coupling the internal and motional dynamics.
In Eq.~\eqref{Newton}, the conservative trap and Coulomb forces are determined from their corresponding potentials:
\begin{equation}
\label{cohForces}
\begin{aligned}
	\bs F^\text{trap}_j & = - \nabla_{\bs r_j} V_\text{trap}; \\
	\bs F^\text{Coul}_j & = - \nabla_{\bs r_j} V_\text{Coul},  \\
\end{aligned}
\end{equation}
and the net laser force on ion $j$ is
\begin{equation}
\label{laserForces}
	\bs F^\text{laser}_j = -i
			\sum_{l,\alpha,\beta} \bs k_l \frac{\Omega_{\alpha \beta}^{(l,j)}}{2} \braket{  \hat{\sigma}_{\alpha \beta}^{(j)} }e^{- i  \phi^{(l,j)}_{\alpha \beta}(\bs r_j,t)}
			+ \text{c.c.}.
\end{equation}

We simplify the problem further by fixing the motion to follow only the dynamics of the external potential at a fixed total mechanical energy and calculating the average rate of energy change due to the viscous laser forces in the presence of this motion.
We refer to this approach as the ``power-averaged cooling method for analyzing $\bar n$," or PACMAN.
This effectively averages over the internal and trap timescales and provides a model whose dynamics includes only the cooling timescale.
We expect this to be a good approximation in the typical case where the laser cooling process occurs on a timescale that is long compared to the trap motion, which is typically harmonic with period $\lesssim 1 \, \mu$s.
Moreover, when the external potential is harmonic, the rate of change of energy can be easily calculated on a mode-by-mode basis.
PACMAN is particularly useful at an intermediate energy scale such that the cooling dynamics differ from the zero-energy limit, but trap and Coulomb nonlinearities are negligible.

For a harmonic potential, in the normal mode basis, the instantaneous laser force affecting each mode is
\begin{equation}
	F^\text{laser}_\mu (t) = \sum_{ij} N_{ij\mu} \, F_{ij}^\text{laser}(t),
\end{equation}
where $F_{ij}$ the the $i^\text{th}$ component of the force on ion $j$.
Here, $N_{i j \mu}$ is the participation of coordinate $i$ of ion $j$ in normal mode $\mu$ determined from the Hessian matrix of $V_\text{trap} + V_\text{Coulomb}$, i.e.,
\begin{equation}
  r_{i j}(t) = \sum_\mu N_{i j \mu} x_\mu(t),
\end{equation}
where mode coordinate $x_\mu$ is associated with a mode of angular frequency $\omega_\mu$ and oscillator length $x_{\mu,0} = \sqrt{1/(m_\mu \omega_\mu)}$, and $m_\mu$ is an effective mass.
The instantaneous, semiclassical rate of energy change for each mode is then
\begin{equation}
\label{powerE}
	\frac{dE_\mu}{dt}(t) = F^\text{laser}_\mu(t) \, v_\mu(t),
\end{equation}
in which $v_\mu = d x_\mu/dt$ is the mode velocity.
Motivated by the correspondence principle, we can relate the total mechanical energy of each mode $E_\mu$ to that of a coherent state $\ket{\alpha}$ by parametrizing $E_\mu$ as
\begin{equation}
	E_\mu = \braket{\alpha_\mu | \hat a_\mu^\dag \hat a_\mu| \alpha_\mu} \omega_\mu = \mathscr n_\mu \omega_\mu,
\end{equation}
wherein $\mathscr n_\mu \equiv  \left|\alpha_\mu\right|^2$ denotes the average occupancy of a coherent state.
The mode position $x_\mu$ and velocity $v_\mu$ then satisfy
\begin{align}
  \label{xv}
  \begin{split}
    x_\mu(\mathscr n_\mu,t) &= \sqrt{\frac{2 \mathscr n_\mu}{m_\mu \omega_\mu}}\cos(\omega_\mu t + \theta_\mu), \\
    v_\mu(\mathscr n_\mu,t) &= -\sqrt{\frac{2 \mathscr n_\mu \omega_\mu}{m_\mu}}\sin(\omega_\mu t + \theta_\mu),
  \end{split}
\end{align}
in which $\theta_\mu$ is the initial secular phase.

Explicitly notating the dependence on mode occupancies, we can rewrite Eq.~\eqref{powerE} as 
\begin{equation}
\label{powerN}
	\omega_\mu \frac{d\mathscr n_\mu}{dt}(\bs{\mathscr n},t) = F^\text{laser}_\mu(\bs{\mathscr n},t) \, v_\mu(\mathscr n_\mu,t),
\end{equation}
where we use the notation $\bs{\mathscr n} = \mathscr n_1, \mathscr n_2, \ldots$ to denote dependence on the average occupancies of all modes. 
Importantly, we find that the cooling dynamics of mode $\mu$ depends on the energy contained in the other modes~\cite{rabl2010}, an effect we will explore further in Section~\ref{multiMode}.

What we are more interested in is the average rate of energy removal given a particular energy configuration of the ion crystal, which is
\begin{equation}
\label{dndt}
	R_\mu(\bs{\mathscr n}) \equiv \lim_{\tau \rightarrow \infty} \frac{1}{\tau} \int_0^\tau   \frac{d\mathscr n_\mu}{dt}(\bs{\mathscr n},t)   \, dt.
\end{equation}
In practice, the averaging time $\tau$ must be long compared to the secular dynamics, i.e., $\tau \gg \frac{2\pi}{\omega_1}, \frac{2 \pi}{\omega_2}, \ldots$, but can be much shorter than the total cooling time.
Although we do not denote so in Eq.~\eqref{dndt} for notational simplicity, an average over the initial secular phases $\theta_\mu$ must also be performed to appropriately sample the classical phase space.

We apply one more transformation to make our comparisons to other cooling theories more transparent.
The term ``cooling rate" typically refers to the decay constant $\mathcal{W}_\mu$ in the exponential cooling equation~\cite{Cirac1992}
\begin{equation}
\label{LDexponentialcooling}
	\frac{d \bar n_\mu}{dt} = - \mathcal{W}_\mu \bar n_\mu + A^+_\mu,
\end{equation}
wherein the term $A^+_\mu$ incorporates heating from blue sideband transitions and diffusion due to recoil from spontaneous emission.
We do not track such effects with PACMAN, so we omit $A^+_\mu$ to define the semiclassical cooling rate for each mode to be
\begin{equation}
\label{semiclassicalCoolingRate}
	W_{\text{SC}, \mu}(\bs{\mathscr n}) \equiv -  \frac{R_\mu(\bs{\mathscr n})}{\mathscr n_\mu}.
\end{equation}
Expressing the cooling dynamics in terms of an energy-dependent cooling rate emphasizes the degradation of exponential cooling beyond the Lamb-Dicke regime that we will discuss in later sections.

\section{Comparison to quantum models}
\label{quantumComparison}

In this section, we compare PACMAN to two well-studied models which treat the motional dynamics quantumly, and therefore serve to benchmark its accuracy.
One is a simulation of the internal and motional dynamics on the composite Hilbert space, referred to as ``fully quantum" or ``FQ," and the other is a Lamb-Dicke regime model that takes advantage of the condition 
\begin{equation}
\label{LDcondition}
	\eta \sqrt{\braket{(\hat{a} + \hat{a}^\dag)^2}} \ll 1,
\end{equation}
referred to as ``LD."
Both models use a quantum master equation in the harmonic approximation.
The FQ model treats the ion-laser interaction $\mathcal L_\text{int} \hat \rho$ to all orders in $\eta$ but expands the dissipator $\mathcal L^\mathcal{D} \hat \rho$ to second order, whereas the LD model expands both $\mathcal L_\text{int} \hat \rho$ and $\mathcal L^\mathcal{D} \hat \rho$ to second order in $\eta$ and couples in the mechanical effects of the laser on the ion to lowest order in perturbation theory~\cite{Cirac1992}.

\begin{figure}
	\includegraphics[width=0.8\linewidth]{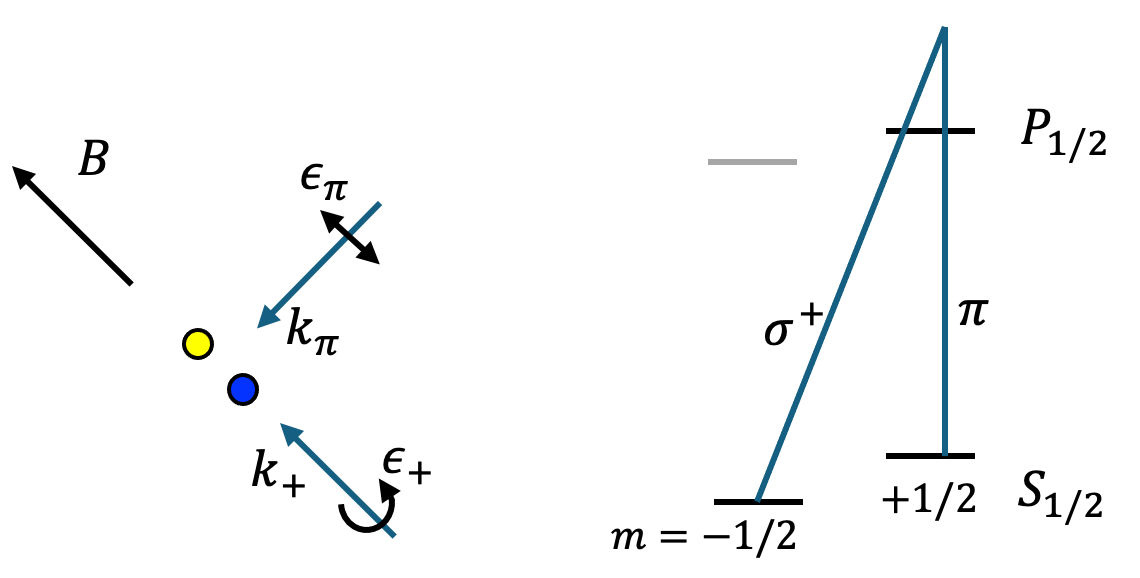}
	\caption{Left: $\bs B$-field and laser geometry. Right: $\Lambda$-system embedded in the $6S_{1/2}$ and $6P_{1/2}$ manifolds of ${}^{138}$Ba$^+$.}
	\label{fig:simpleEIT}
 \end{figure}

For simplicity, we focus our comparison to EIT cooling a single mode with an internal $\Lambda$-system.
We choose the axial stretch mode of a two-ion crystal consisting of ${}^{138}$Ba$^+$ and ${}^{171}$Yb$^+$ ions (BY crystal) with frequency $\omega = 2\pi \, 1.84 \,$MHz and associated two-photon Lamb-Dicke parameter $\eta = 0.044$.
The cooling takes place in a simplified subspace embedded in the $6S_{1/2}$ and $6P_{1/2}$ manifolds of barium, as shown in Fig.~\ref{fig:simpleEIT}.
In our simulations, we artificially ignore the off-resonant single-photon $\pi$-transition between the $m=-1/2$ states, as well as leakage to the $5D_{3/2}$ manifold, restricting the dynamics to within the $\Lambda$-system.
(These effects will be included in later sections.)
The laser detunings are both fixed to $\Delta = 2\pi \, 25 \,$MHz so that the carrier transition is nulled by the EIT resonance.
The Rabi rates $\Omega_+, \Omega_\pi$ are chosen by a particular optimization of the cooling rates and steady-state average occupancies $\bar n_\text{ss}$ of all crystal modes as predicted by the LD model.
 
\begin{figure}
	\input{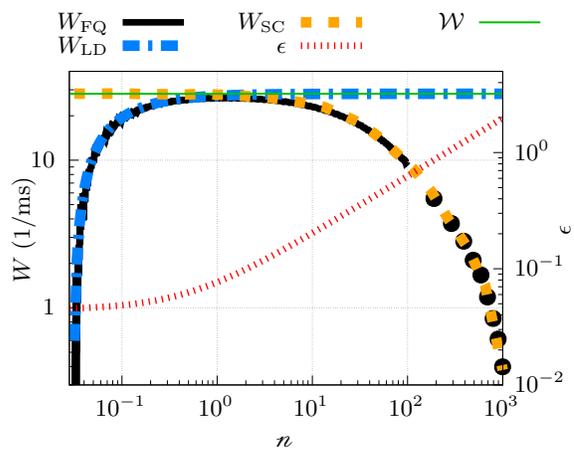}
	\caption{Cooling rates $W$ as a function of coherent state average occupancy $\mathscr n$  for the axial stretch mode of a BY crystal calculated from the semiclassical (SC) PACMAN model [orange, dashed, Eq.~\eqref{semiclassicalCoolingRate}], Lamb-Dicke (LD) regime model [blue, dot-dashed, Eq.~\eqref{WLD}], and fully-quantum (FQ) model [black, solid and points, Eq.~\eqref{Wquantum}]. The FQ prediction agrees well with the SC model when  $\epsilon > 0.1$ [red, dotted, Eq.~\eqref{epsilon}]. Parameters are: $\omega/2\pi = 1.84 \, \text{MHz}, \, \eta = 0.044, \, \Delta/2\pi = 25 \, \text{MHz}, \, \Omega_+/2\pi =\Omega_\pi/2\pi =  11. \, \text{MHz}$, and $\Gamma_+/2\pi = 2\Gamma_\pi/2\pi = 10.1 \, \text{MHz}.$}
	\label{fig:ratesvsn}
 \end{figure}

 The predicted cooling rate $W(\mathscr n)$ for each cooling model is presented in Fig.~\ref{fig:ratesvsn} and were calculated as follows. 
For PACMAN, we used an averaging time $\tau$ of 10 $\mu$s [see Eq.~\eqref{semiclassicalCoolingRate}].
To allow for the internal dynamics to first reach a dynamic steady state, we evolved the system for 5 $\mu$s before averaging, yielding a total cooling simulation time of 15 $\mu$s.
Because there are no other modes in consideration, no relative secular phase averaging was required.

As with the semiclassical definition, we define the quantum cooling rates as $W(\bar n) \equiv - (d \bar n/dt)/\bar n$.
Because the LD model predicts exponential cooling, as in Eq.~\eqref{LDexponentialcooling}, we find 
\begin{equation}
\label{WLD}
	W_\text{LD}(\bar n) = \mathcal{W} - \frac{A^+}{\bar n}.
\end{equation}
Notice that residual heating yields a negative cooling rate below the steady-state occupancy $\bar n_\text{ss} = A^+/\mathcal{W}$.

For the fully-quantum model, we calculated the cooling rate from the quantum master equation at each time $t$:
\begin{equation}
\label{Wquantum}
	W_\text{FQ}(\bar n) = - \frac{1}{\tr{\hat n \hat \rho}} \tr{\hat n \, \frac{d \hat \rho}{dt}}.
\end{equation}
This approach allows us to parametrically calculate the cooling rate $W_\text{FQ}[\bar n(t)]$ shown as the black line in Fig.~\ref{fig:ratesvsn}.
While the LD prediction for the cooling rate is not sensitive to the initial motional distribution $\hat \rho_m \equiv \text{Tr}_I\left[\hat \rho\right]$, we must consider what initial distribution to use in the FQ model.
Because the classical motion is harmonic, and the secular phase of the harmonic motion is irrelevant to the cooling dynamics (and is often experimentally uncontrollable), we initialized $\hat \rho_m$ in a phase-averaged coherent (PAC) state~\cite{allevi2013}:
\begin{equation}
\label{PAC}
	\hat \rho_\text{PAC}(\mathscr n) = \frac{1}{2 \pi} \int_0^{2\pi} \hspace{-0.6em} \ket{\alpha}\bra{\alpha} d \theta = e^{-\mathscr n} \sum_{n=0}^\infty \frac{{\mathscr n}^n}{n!} \ket{n}\bra{n},
\end{equation}
in which we have used the polar representation $\alpha = \sqrt{\mathscr n} e^{i \theta}$.
The initial quantum state was chosen to be
\begin{equation}
\label{quantumInit}
	\hat \rho(t=0) = \ket{\psi_\text{dark}} \bra{\psi_\text{dark}} \otimes \hat \rho_\text{PAC}(\mathscr n_0),
\end{equation}
where $\ket{\psi_\text{dark}}$ is the EIT dark state~\cite{morigi2000}
\begin{equation}
	\ket{\psi_\text{dark}} = \frac{1}{\sqrt{\Omega_1^2 + \Omega_2^2}}
		\left(
		\Omega_1 \ket{2} - \Omega_2 \ket{1}
		\right).
\end{equation}

At high energies ($ \bar n > 100$), the FQ simulation becomes too numerically intensive to simulate cooling over long time scales.
To circumvent this issue, we approximated the cooling rate of a PAC state with initial average occupancy $\mathscr n_0$ by simulating the cooling within a window of Fock states centered on $\mathscr n_0$ and calculating $W_\text{FQ}(\bar n)$ after 10 $\mu$s.
This allows for the internal dynamics to equilibrate, but does not significantly change the motional distribution, i.e., $\bar n(t) \approx \mathscr n(t)$.
Cooling rates for the FQ model calculated in this way are included as black points in Fig.~\ref{fig:ratesvsn}.

To quantify the regimes of validity for the LD and PACMAN models, we have also plotted
\begin{equation} 
\label{epsilon}
	\epsilon \equiv \eta \sqrt{\braket{(\hat{a} + \hat{a}^\dag)^2}_\text{PAC}} = \eta \sqrt{2 \mathscr n + 1},
\end{equation}
which was calculated from Eq.~\eqref{PAC}.
We find that the LD (blue, dot-dashed) and FQ (black, solid and points) models agree for $\epsilon < 0.1$, and that the PACMAN (orange, dashed) and FQ models agree for $\epsilon > 0.1$.
The steep drop of the LD and FQ curves on the left side of the plot is due to the residual heating from spontaneous recoil, which is not included in PACMAN.
However, $W_\text{SC}$ saturates to the (fixed) exponential quantum cooling rate $\mathcal{W}$ in Eq.~\eqref{WLD} (green, thin, solid) in the limit $\mathscr n \rightarrow 0$.

Interestingly, the PACMAN and FQ cooling rates substantially degrade from the LD prediction for $\mathscr n > 10$, indicating reduced cooling performance at experimentally relevant energy scales.
At high energies, this can be partially attributed to the reduction in size of the Franck-Condon factors
\begin{equation}
|\braket{n'|e^{i \bs k \cdot \bo{r}}|n}| \approx |J_{|n-n'|}(2 \eta \sqrt{n})| \leq \sqrt{\frac{\pi}{\eta \sqrt{n}}},
\end{equation}
where $J$ is a Bessel function of the first kind~\cite{wineland1979,hu2011,peik1999,rabl2010,highTempOscillator}.
Another source of cooling degradation, which we discuss in Appendix~\ref{captureRange}, is the emergence of significant Doppler heating wherein the effects of blue-detuned laser photon absorption overwhelm the coherent EIT Raman transitions.
In fact, the cooling rate can become negative at high energies, resulting in a runaway heating effect.

Given that the fully-quantum result for $1 < \mathscr n \leq 100$ was calculated from a continuous simulation over which the form of the quantum motional distribution could, in principle, substantially change, it is not obvious that the PACMAN and fully-quantum predictions should agree over this energy range.
We explore the evolution of the quantum motional distribution in more detail in Appendix~\ref{changes} to provide insight into how long the predicted semiclassical cooling rates accurately reflect the true cooling dynamics.

\section{Comparison to experiment}
\label{expComparison}

We further benchmark the accuracy of PACMAN by comparing its predictions to an experiment wherein a single mode of a pre-cooled BY crystal is coherently excited to a high energy and then cooled near to its motional ground state.
This comparison also demonstrates the utility of PACMAN in realistic scenarios in which many internal levels can affect the cooling process.

\begin{figure}
	\includegraphics[width=\linewidth]{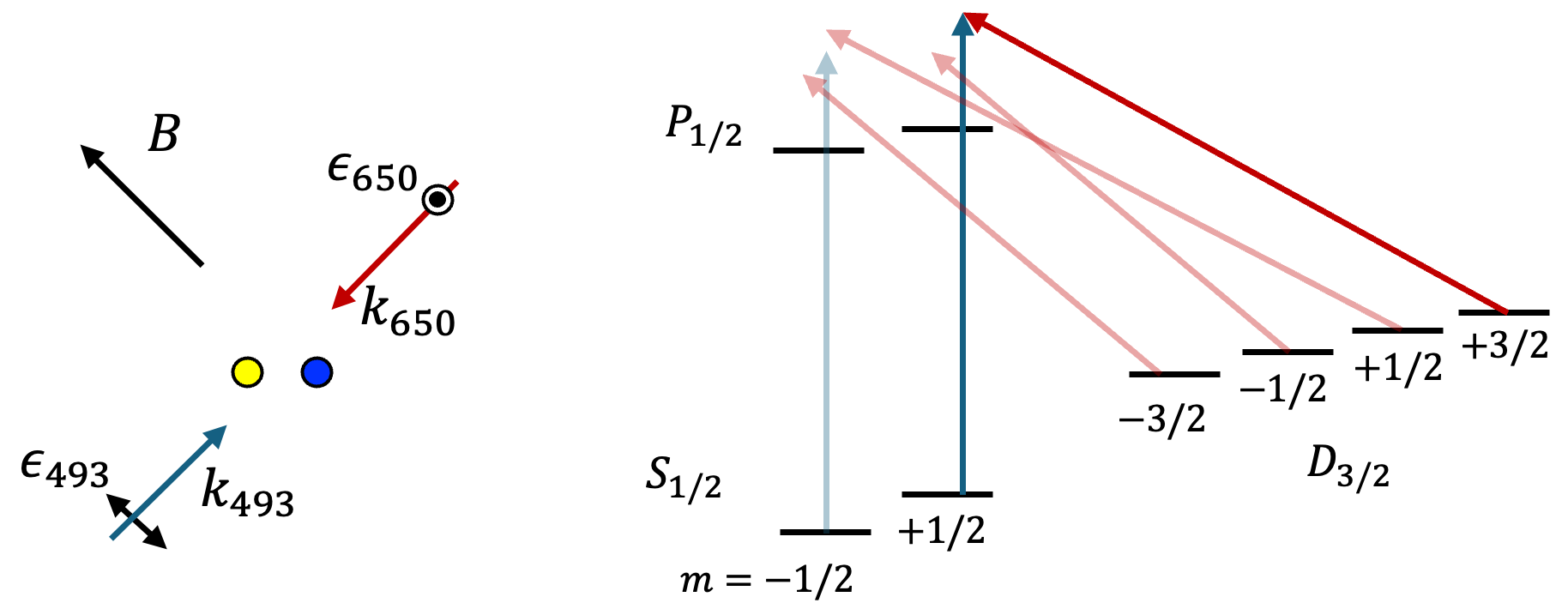}
	\caption{Left: $\bs B$-field and laser geometry. Right: $\Lambda$-subsystem embedded in the $6S_{1/2}$, $5D_{3/2}$ and $6P_{1/2}$ manifolds of ${}^{138}$Ba$^+$. The dark arrows label the resonant EIT transition, and the light arrows label the other possible laser transitions.}
	\label{fig:marmotLevels}
 \end{figure}

\begin{figure}
	\input{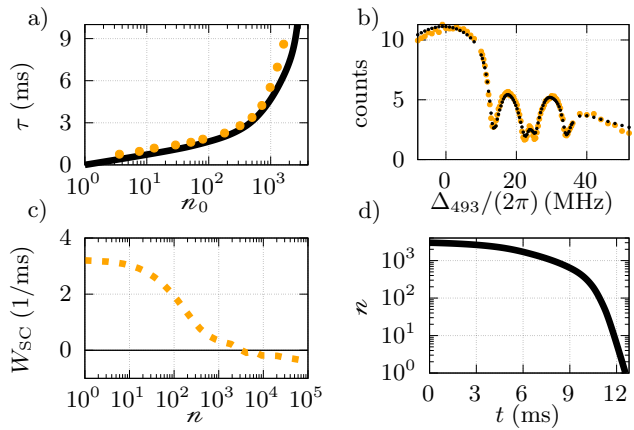}
	\caption{Comparison of the PACMAN model to an EIT cooling experiment using the axial COM mode of a BY crystal. Parameters are: $|\bs B| =5.27 \, \text{G}, \, \omega = 2\pi \, 1.05 \, \text{MHz}, \, \eta = 0.059, \, \Delta_{493}/2\pi = 25.35 \, \text{MHz}, \, \Delta_{650}/2\pi = 23.94 \, \text{MHz}, \, I_{493} = 0.784 \, \text{mW/mm}^2, \, I_{650} = 0.906 \, \text{mW/mm}^2, \, \gamma_{493}/2\pi = 0 \, \text{kHz}$, and $\gamma_{650}/2\pi = 146 \, \text{kHz}$. The laser detunings are measured relative to the centers of the manifolds. a) Time $\tau$ to cool to a thermal occupancy $\bar n_{th} =1$ vs. initial coherent excitation $\mathscr n_0$ as predicted by the experiment (orange, points) and PACMAN (black, solid). b) Fluorescence measurement vs. detuning of the 493 nm laser $\Delta_{493}$. The laser parameters used in the simulation are extracted from a fit (black points) to the experimental values (orange points). c) Semiclassical cooling rate $W_\text{SC}(\mathscr n)$ (orange, dashed) using PACMAN. We observe an EIT capture range (black, solid) at $\mathscr n \approx 3000$. d) Average coherent state mode occupancy $\mathscr n$ vs. time $t$ during EIT cooling calculated by interpolating $W_\text{SC}$. }
	\label{fig:MarmotEIT}
 \end{figure}

The utilized scheme is a novel EIT cooling method we propose which uses near-detuned lasers to induce fast cooling over a broad range of motional mode frequencies to low temperatures.
Due to leakage to the $5D_{3/2}$ manifold, we must consider a different internal level coupling scheme beyond the simple $\Lambda$-system used in Section~\ref{quantumComparison}.
Instead of optically repumping leaked population from $5D_{3/2}$ into $6S_{1/2}$, we decide to use one of its Zeeman sublevels in a $\Lambda$-subsystem, as shown in Fig.~\ref{fig:marmotLevels}.
Specifically, we target the EIT dark state comprised of the states $\ket{6S_{1/2},m_F=1/2}$ and $\ket{5D_{3/2},m_F=3/2}$, which corresponds to the third fluorescence dip from the left in Fig.~\ref{fig:MarmotEIT}(b).
EIT dark resonances between other Zeeman sublevels are avoided by introducing Zeeman shifts with a uniform $|\bs B| = 5.27$ G magnetic field.
The EIT lasers also serve to optically repump population leaked outside of the EIT subspace.
The cooling is again implemented with a pair of lasers, one inducing pure $\pi$-transitions (linear polarization along the $\bs B$-field) between $S_{1/2}$ and $P_{1/2}$ with a wavelength of 493 nm, and the other inducing $\sigma^\pm$-transitions (linear polarization perpendicular to the $\bs B$-field) between $D_{3/2}$ and $P_{1/2}$ with a wavelength of 650 nm.
The crystal axis is aligned such that the difference wavevector $\Delta \bs k \equiv \bs k_{493} - \bs k_{650}$ has equal projections onto all principal axes to permit all-mode cooling.
We use the same trap frequencies as in Section~\ref{quantumComparison}, but now focus on the axial center-of-mass (COM) mode since it is easiest to excite coherently without affecting other modes.
For this setup, the axial COM mode has frequency $\omega = 2\pi \, 1.05 \,$MHz and two-photon Lamb-Dicke parameter $\eta = 0.059$.

The experimental cooling procedure is as follows. 
The cooling rates for the axial COM and radial stretch modes were optimized experimentally by starting at Doppler temperature and varying the laser intensities and detunings in the region of the EIT dark state. 
The optimal experimental laser parameters were chosen as a compromise to maximize the cooling rate for all modes. 
To measure the cooling rate from a high-energy state, all crystal modes were first cooled to the ground state with a combination of EIT and resolved-sideband cooling. 
The crystal was then coherently excited by a calibrated axial well displacement to deterministically excite the axial COM mode to the average occupancy $\mathscr n_0$. 
The mode was EIT cooled for a time $t$ and (assuming the motional distribution was thermalized by the cooling process) fit to a thermal distribution to extract an average occupancy $\bar n_\text{th}(t)$.
The thermal fit consisted of measuring Rabi flops between the ytterbium clock states $\ket{\downarrow} \equiv \ket{{}^2S_{1/2},F=0,m=0}$ and $\ket{\uparrow} \equiv \ket{{}^2S_{1/2},F=1,m=0}$ on the carrier, red, and blue sideband transitions~\cite{meekhof1996}.
The cooling time $t$ was varied over five values such that $0.5 < \bar n_\text{th}(t) < 20$ and then fit to the exponential decay function
\begin{equation}
	\bar n_{th}(t) = (\bar n_0 - \bar n_{ss}) e^{-\mathcal W t} + \bar n_{ss}.
\end{equation}
The exponential fit was then used to determine how long it took to re-cool from the coherent state to the thermal average occupancy $\bar n_{th} = 1$, i.e., the re-cooling time $\tau$ that satisfies $\tau = t(\bar n_{th} =1)$.
(We chose the final occupancy $\bar n_{th} = 1$ so that the system underwent significant cooling, but the temperature remained sufficiently above its steady-state value, which is not captured by the semiclassical model.)
The experiment was repeated with different initial coherent excitations $\mathscr n_0$ and cooling times. 
The results of this procedure are shown as orange points in Fig.~\ref{fig:MarmotEIT}(a). 

To compare to experiment, PACMAN was used to predict $\tau$ in the following manner.
To account for uncertainties in the laser parameters at the position of the ion crystal, some of the model’s parameters were determined by fitting to an experimental measurement of the EIT lineshape, which is shown as orange points in Fig.~\ref{fig:MarmotEIT}(b).
The lineshape was obtained by scanning the detuning of the $493$ laser $\Delta_{493}$ over the detuning of the 650 laser $\Delta_{650}$ at fixed laser power and measuring the fluorescence rate with a photodetector.
The ions were Doppler cooled before each fluorescence measurement.
A fit to the experimental data was then done by calculating the steady-state fluorescence rate and varying the simulated laser intensities $I_{493}, I_{650}$, detunings, and linewidths $\gamma_{493}, \gamma_{650}$ (black, solid).
To account for Doppler broadening, we included thermal motion at the experimentally-measured Doppler temperatures in the fluorescence rate calculation.
We used the fitted laser parameters to calculate a semiclassical cooling rate $W_\text{SC}(\mathscr n)$ [see Eq.~\eqref{semiclassicalCoolingRate}] for various $\mathscr n$ with PACMAN (orange, dashed); the results are displayed in Fig.~\ref{fig:MarmotEIT}(c).
To allow for the internal dynamics to first reach a dynamic steady state, we first simulated the system for 5 $\mu$s.
We then averaged the simulation data over the following 20 $\mu$s, yielding a total cooling simulation time of 25 $\mu$s.
The semiclassical cooling rate $W_\text{SC}(\mathscr  n)$ was interpolated to calculate $\mathscr n(t)$ with the initial condition $\mathscr n(0) = 3000$ [Fig.~\ref{fig:MarmotEIT}(d)].
The simulated $\mathscr n (t)$ from PACMAN was then used to predict $\tau$ by calculating the time taken to evolve from $\mathscr n_0$ to $\mathscr n = 1$, which is included in Fig.~\ref{fig:MarmotEIT}(a) (black, solid).

We now discuss the results of the cooling experiment and PACMAN.
Most importantly, as shown in Fig.~\ref{fig:MarmotEIT}(a), we observe strong agreement between experiment and theory for the re-cooling time $\tau$ over a broad energy range.
In the experiment, we were unable to EIT cool the mode back to the ground state for excitations $\mathscr n_0 \geq 2000$ in the experiment.
This is roughly consistent with the theoretical prediction of the EIT capture range of $\mathscr n_\text{cap} \approx 3000$, which can be seen in Fig.~\ref{fig:MarmotEIT}(c) as $W_\text{SC}(\mathscr n_\text{cap}) = 0$.
The degradation of exponential cooling can also be observed in Fig.~\ref{fig:MarmotEIT}(c) for $\mathscr n \geq 10$. 
These results further demonstrate the accuracy and utility of PACMAN and provide more insight into the evolution of the quantum motional distribution throughout the cooling process, as discussed in Appendix~\ref{changes}.

The EIT heating effect has been observed in similar studies~\cite{highTempOscillator,kepesidis2013}, wherein it is attributed to motional decoherence of the Raman transitions such that the dynamics are instead dominated by Doppler heating.
We provide evidence for this claim in Appendix~\ref{captureRange}, wherein we calculate the EIT capture range estimate
\begin{equation}
	\mathscr n_\text{cap} \approx \frac{\Omega^2 - 2\omega^2}{2 \eta^2 \omega^2}.
\end{equation}
With the additional requirement of fixing the Raman Rabi frequency $\Omega_R \propto \Omega^2/\Delta$, we find that the capture range can be extended to arbitrarily large values at the cost of increased laser power and lower overall cooling rates.

\begin{figure*}
	\input{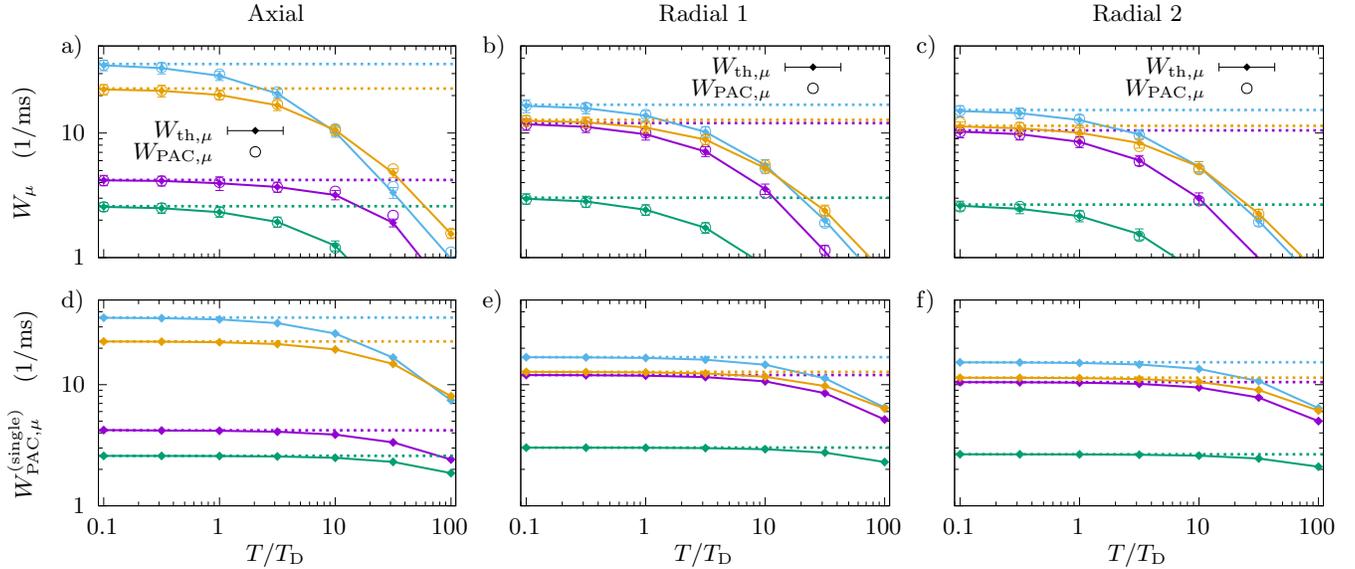}
	\caption{Cooling rates $W_\mu$ for the 12 modes of a YBBY crystal with various motional distributions under the cooling scheme depicted in Fig.~\ref{fig:marmotLevels} as a function of temperature $T$, in units of the Doppler temperature $T_\text{D} = \hbar \Gamma/2$. We have grouped the data by principal axis. At low energies, all PACMAN predictions agree with the exponential cooling rates $\mathcal W_\mu$ predicted by the LD model (dashed). (a-c)  Comparison of the cooling rate for the 12-mode thermal state [Eq.~\eqref{Wthermal}, solid with diamonds] and the 12-mode PAC state [Eq.~\eqref{Wall}, circles], with all modes containing the energy corresponding to the temperature $T$. (d-f) Cooling rate for the 12-mode PAC state with only one mode containing the energy corresponding to the temperature $T$ [Eq.~\eqref{Wone}]. Parameters are:  Yb trap frequencies (axial, radial 1, radial 2) = $(1.00, 2.72, 2.82) \, 2\pi \, \text{MHz}, \, B = 5.27 \, \text{G}, \, \Delta_{493}/2\pi = 25.2 \, \text{MHz}, \, \Delta_{650}/2\pi = 23.7 \, \text{MHz}, I_{493} = 0.46 \, \text{mW/mm}^2, \, I_{650} = 1.51 \, \text{mW/mm}^2, \, \gamma_{493}/2\pi = 0 \, \text{kHz}$, and $\gamma_{650}/2\pi = 146 \, \text{kHz}$.}
	\label{fig:multimode}
 \end{figure*}

\section{Cooling dynamics for any phase-averaged quantum motional distribution}
\label{instantaneousRates}

The cooling results we have presented thus far have focused on single-mode coherent states.
However, it is often of interest to calculate the cooling dynamics for an $N$-mode quantum motional distribution where each mode $\mu$ may not be in a coherent state.
We focus on phase-averaged states, a class of states which are diagonal in the Fock basis, which is usually sufficient for laser-cooled systems.
Such states can be represented as
\begin{equation}
\label{Prep_symm}
	\hat \rho = \pi^N \int d^N \bs{\mathscr n} \, \hat \rho_\text{PAC}^{\otimes N}(\bs{\mathscr n}) \, P (\bs{\mathscr n}),
\end{equation}
wherein we have introduced the $N$-mode PAC state [see Eq.\eqref{PAC}], as well as the $P$-function from the Glauber-Sudarshan $P$ representation~\cite{carmichaelvol1}.

We show in Appendix~\ref{instantaneousRatesAppendix} that the moments
\begin{equation}
	\braket{\hat n_\mu^k} =  \tr{\hat \rho \, \hat n_\mu^k}
\end{equation}
for phase-averaged states, where $k = 1,2,\ldots$, satisfy
\begin{equation}
\label{dndtP}
\begin{aligned}
	R^{(k)}_\mu & \equiv
	\frac{d}{dt}\braket{\hat n_\mu^k}
	\\
	= -& k \pi^N \int d^N \bs{\mathscr n} \,  \mathscr n_\mu^k \, W_{\text{SC}, \mu}(\bs{\mathscr n}) \, P(\bs{\mathscr n}).
\end{aligned}
\end{equation}
The rates in Eq.~\eqref{dndtP} can be used to benchmark the early performance of a cooling scheme when the pre-cooled motional distribution is known.

The approach presented in Section~\ref{semiRates} fits into this generalized formalism through the $N$-mode PAC state $P$-function~\cite{allevi2013}
\begin{equation}
\label{PPAC}
	P_\text{PAC}(\bs{\mathscr n}, \bs{\mathscr n}') = \prod_\mu \frac{1}{\pi}\,\delta(\mathscr n_\mu' - \mathscr n_\mu),
\end{equation}
in which $\delta$ is the Dirac delta function. 
Substituting Eq.~\eqref{PPAC} into Eq.~\eqref{dndtP} with $k=1$ and integrating over $\bs{\mathscr n}'$ results in
\begin{equation}
\label{dndtPAC}
	R^{(1)}_{\text{PAC},\mu} =  - W_{\text{SC}, \mu} (\bs{\mathscr n}) \, \mathscr n_\mu,
\end{equation}
which is identical to Eq.~\eqref{semiclassicalCoolingRate}.

A pragmatic example is the case when all modes are in thermal states with average occupancies $\bs{\bar n}_\text{th} = \bar n_{\text{th},1}, \ldots, \bar n_{\text{th},N}$.
The $N$-mode thermal $P$-function is
\begin{equation}
\label{Pthermal}
	P_\text{th}(\bs{\bar n}_\text{th},\bs{\mathscr n}) = \prod_\mu \frac{e^{-\mathscr n_\mu/ \bar n_{\text{th},\mu}}}{\pi \bar n_{\text{th},\mu}},
\end{equation}
so that
\begin{equation}
\label{dndt_thermal}
	R^{(1)}_{\text{th},\mu} = - \int d^N \bs{\mathscr n} \, \mathscr n_\mu \, W_{\text{SC}, \mu}(\bs{\mathscr n}) \, P_\text{th}(\bs{\bar n}_\text{th},\bs{\mathscr n}).
\end{equation}
Equation~\eqref{dndt_thermal} is particularly useful when analyzing ground-state cooling methods for modes that have been pre-cooled to a thermal state with Doppler cooling, which we discuss in the next section.
Although it technically requires the calculation of cooling dynamics over an $N$-dimensional space, we can drastically simplify the computational requirements through Monte Carlo sampling or by considering a simpler distribution, as demonstrated in the next section.

\section{Multi-mode cooling}
\label{multiMode}

In this section, we demonstrate how PACMAN can be used to predict multi-mode cooling rates.
Further, we use the techniques of Section~\ref{instantaneousRates} to determine cooling rates for various motional distributions.
As a high-dimensional example, we demonstrate effective cooling of a YBBY crystal, which contains $N=12$ modes, using the barium EIT cooling scheme presented in Section~\ref{expComparison}. 

By the spatial symmetry of the crystal, the coolant ions undergo similar dynamics in the harmonic approximation.
This allows us to simplify the calculation by tracking the internal evolution of a single coolant ion and doubling the resulting cooling rate.
This simplification ignores any entanglement between the coolant ions, which we assume to be significant only at low temperatures~\cite{homeThesis}.

The semiclassical cooling rate predictions are displayed in Fig.~\ref{fig:multimode}.
We chose the laser parameters by optimizing the exponential cooling rates and steady-state average occupancies with the LD model, the former being displayed as dashed lines.
The LD model predicts very low steady-state average occupancies $\bar n_{\text{ss},\mu} \leq 0.15$ for all modes except the axial COM mode, which relaxes to $\bar n_\text{ss} =0.38$.
Eight of the 12 modes, which span a 1 MHz frequency range, exhibit fast cooling time constants $\tau_\mu= 1/\mathcal{W_\mu} < 100 \, \mu$s, demonstrating the broadband nature of the cooling method.

We first consider the case when the crystal modes have mutually thermalized, so their motional distributions are characterized by
\begin{equation}
\label{temperature}
	\frac{1}{2} k_B T = \hbar  \bar n_{\text{th}, \mu} \omega_\mu.
\end{equation}
(In particular, crystals which have first been Doppler cooled will begin the EIT cooling on the order of the Doppler temperature $k_B T_\text{D} = \hbar\Gamma/2$~\cite{leibfried2003}.)
The motional distribution is then the 12-mode thermal $P$-function, $P_\text{th}$ [as in Eq.~\eqref{Pthermal}], with the thermal average 
occupancies determined by Eq.~\eqref{temperature}.
From Eq.~\eqref{dndt_thermal}, the thermal cooling rates are
\begin{equation}
\label{Wthermal}
	W_{\text{th},\mu} = -\frac{R^{(1)}_{\text{th},\mu}}{\bar n_{\text{th},\mu}},
\end{equation}
which are displayed in Figs.~\ref{fig:multimode}(a-c) (solid with diamonds).
For each temperature $T$, we drew 100 random samples from $P_\text{th}$ to sample the 12-dimensional space, and we averaged each cooling rate over at least 16 PACMAN trajectories with randomized secular phases.
At low temperatures $T \ll T_\text{D}$, we find that all modes exhibit exponential cooling at the constant rate $\mathcal W_\mu$ determined by the LD model.
At higher temperatures $T \gg T_\text{D}$, the cooling rates of all modes significantly degrade from their low-temperature values.
This result demonstrates the need to sufficiently pre-cool the distribution to realize fast low-temperature EIT cooling rates.
However, it can be difficult to cool all modes of a non-homogeneous crystal with one sympathetic species since some modes may have small LD parameters; the energy remaining in these poorly-cooled modes may significantly affect the cooling rates of otherwise fast-cooling modes~\cite{rabl2010}.

To emphasize the dependence of the cooling rate on the underlying motional distribution, we also consider the 12-mode PAC state $P$-function, $P_\text{PAC}$ [as in Eq.~\eqref{PPAC}].
As an approximation to the thermal case, we first calculate cooling rates when all modes contain the energy corresponding to the temperature $T$,
\begin{equation}
\label{PPACall}
	P_\text{PAC} = \prod_\mu \frac{1}{\pi}\,\delta\left(\mathscr n_\mu - \bar n_{\text{th},\mu} \right),
\end{equation}
which, from Eq.~\eqref{dndtPAC}, corresponds to the cooling rate
\begin{equation}
\label{Wall}
	W_{\text{PAC},\mu} =  W_{\text{SC}, \mu} \left(\bs{\mathscr n} = \bs{\bar n}_\text{th}\right).
\end{equation}
The results are included as circles in Figs.~\ref{fig:multimode}(a-c).
We find that this distribution yields similar predictions to $W_{\text{th},\mu}$ while requiring drastically fewer numerical samples, as the distribution is one-dimensional.

Next, to compare to a single-mode approach~\cite{highTempOscillator}, we calculate cooling rates when only mode $\mu$ contains the energy corresponding to the temperature $T$ in Eq.~\eqref{temperature}, and all other modes are at zero energy:
\begin{equation}
\label{PPACone}
	P_{\text{PAC},\mu}^{(\text{single})} = 
	\frac{1}{\pi}\,\delta\left(\mathscr n_\mu - \bar n_{\text{th},\mu}\right)
	\prod_{\nu \neq \mu} \frac{1}{\pi}\,\delta(\mathscr n_\nu).
\end{equation}
This distribution corresponds to the cooling rate
\begin{equation}
\label{Wone}
	W^{(\text{single})}_{\text{PAC},\mu} =  W_{\text{SC}, \mu} \left(\mathscr n_\mu = \bar n_{\text{th},\mu}, \mathscr n_{\nu} = 0 \right).
\end{equation}
As shown in Figs.~\ref{fig:multimode}(d-f), we observe much higher cooling rates in the high-energy regime, again demonstrating the strong dependence of cooling rate on the entire motional distribution.

\section{Conclusion}

We have presented a semiclassical model that we use to simulate the laser cooling of multiple trapped-ion crystal modes with many internal levels.
We derived a generalized expression to calculate the average rate of energy change of each mode with a computationally efficient model, and demonstrated its accuracy by comparing its predictions to well-established quantum models and to a cooling experiment.
We found that the resulting cooling rates beyond the Lamb-Dicke regime are strongly dependent on the amount of energy in each mode, and are generally lower than their values in the Lamb-Dicke regime.
We have observed an EIT capture range such that a mode can be cooled at low energies, but heated at high energies, indicating the need for sufficient pre-cooling to Doppler temperatures.
We have also shown how the semiclassical predictions can be applied to a more general class of motional distributions, such as thermal states.

Aside from the suppression of cooling rates at high energies due to smaller Franck-Condon factors, it would be interesting to investigate if multi-mode cooling is limited by the rate at which entropy can be removed by spontaneous emission~\cite{enk1992,bartolottaEntropy}.
Although we performed our cooling optimizations with the Lamb-Dicke model (zero-temperature limit), it would be more practical to use the semiclassical model to perform a time-dependent, interemediate-temperature cooling optimization.
For certain cooling methods, our model can be made further computationally efficient by adiabatic elimination of excited internal levels~\cite{morigi1999,reiter2012}.
Our methods can also be generalized to include beyond-harmonic effects, such as micromotion and nonlinear mode couplings~\cite{crossKerr}.
We expect that our methods can be used to motivate cooling methods in experimentally relevant parameter regimes.

\section{Acknowledgements}

We thank J. Gaebler, C.H. Baldwin, R.B. Hutson, M. Schecter, P. Siegfried, J. Dominy, and R.T. Sutherland for stimulating discussions.

\bibliographystyle{apsrev4-2}

\begin{thebibliography}{37}%
\makeatletter
\providecommand \@ifxundefined [1]{%
 \@ifx{#1\undefined}
}%
\providecommand \@ifnum [1]{%
 \ifnum #1\expandafter \@firstoftwo
 \else \expandafter \@secondoftwo
 \fi
}%
\providecommand \@ifx [1]{%
 \ifx #1\expandafter \@firstoftwo
 \else \expandafter \@secondoftwo
 \fi
}%
\providecommand \natexlab [1]{#1}%
\providecommand \enquote  [1]{``#1''}%
\providecommand \bibnamefont  [1]{#1}%
\providecommand \bibfnamefont [1]{#1}%
\providecommand \citenamefont [1]{#1}%
\providecommand \href@noop [0]{\@secondoftwo}%
\providecommand \href [0]{\begingroup \@sanitize@url \@href}%
\providecommand \@href[1]{\@@startlink{#1}\@@href}%
\providecommand \@@href[1]{\endgroup#1\@@endlink}%
\providecommand \@sanitize@url [0]{\catcode `\\12\catcode `\$12\catcode
  `\&12\catcode `\#12\catcode `\^12\catcode `\_12\catcode `\%12\relax}%
\providecommand \@@startlink[1]{}%
\providecommand \@@endlink[0]{}%
\providecommand \url  [0]{\begingroup\@sanitize@url \@url }%
\providecommand \@url [1]{\endgroup\@href {#1}{\urlprefix }}%
\providecommand \urlprefix  [0]{URL }%
\providecommand \Eprint [0]{\href }%
\providecommand \doibase [0]{https://doi.org/}%
\providecommand \selectlanguage [0]{\@gobble}%
\providecommand \bibinfo  [0]{\@secondoftwo}%
\providecommand \bibfield  [0]{\@secondoftwo}%
\providecommand \translation [1]{[#1]}%
\providecommand \BibitemOpen [0]{}%
\providecommand \bibitemStop [0]{}%
\providecommand \bibitemNoStop [0]{.\EOS\space}%
\providecommand \EOS [0]{\spacefactor3000\relax}%
\providecommand \BibitemShut  [1]{\csname bibitem#1\endcsname}%
\let\auto@bib@innerbib\@empty
\bibitem [{\citenamefont {Cirac}\ and\ \citenamefont
  {Zoller}(1995)}]{cirac1995}%
  \BibitemOpen
  \bibfield  {author} {\bibinfo {author} {\bibfnamefont {J.~I.}\ \bibnamefont
  {Cirac}}\ and\ \bibinfo {author} {\bibfnamefont {P.}~\bibnamefont {Zoller}},\
  }\href {https://doi.org/10.1103/PhysRevLett.74.4091} {\bibfield  {journal}
  {\bibinfo  {journal} {Phys. Rev. Lett.}\ }\textbf {\bibinfo {volume} {74}},\
  \bibinfo {pages} {4091} (\bibinfo {year} {1995})}\BibitemShut {NoStop}%
\bibitem [{\citenamefont {Brewer}\ \emph {et~al.}(2019)\citenamefont {Brewer},
  \citenamefont {Chen}, \citenamefont {Hankin}, \citenamefont {Clements},
  \citenamefont {Chou}, \citenamefont {Wineland}, \citenamefont {Hume},\ and\
  \citenamefont {Leibrandt}}]{brewer2019}%
  \BibitemOpen
  \bibfield  {author} {\bibinfo {author} {\bibfnamefont {S.~M.}\ \bibnamefont
  {Brewer}}, \bibinfo {author} {\bibfnamefont {J.-S.}\ \bibnamefont {Chen}},
  \bibinfo {author} {\bibfnamefont {A.~M.}\ \bibnamefont {Hankin}}, \bibinfo
  {author} {\bibfnamefont {E.~R.}\ \bibnamefont {Clements}}, \bibinfo {author}
  {\bibfnamefont {C.~W.}\ \bibnamefont {Chou}}, \bibinfo {author}
  {\bibfnamefont {D.~J.}\ \bibnamefont {Wineland}}, \bibinfo {author}
  {\bibfnamefont {D.~B.}\ \bibnamefont {Hume}},\ and\ \bibinfo {author}
  {\bibfnamefont {D.~R.}\ \bibnamefont {Leibrandt}},\ }\href
  {https://doi.org/10.1103/PhysRevLett.123.033201} {\bibfield  {journal}
  {\bibinfo  {journal} {Phys. Rev. Lett.}\ }\textbf {\bibinfo {volume} {123}},\
  \bibinfo {pages} {033201} (\bibinfo {year} {2019})}\BibitemShut {NoStop}%
\bibitem [{\citenamefont {Eschner}\ \emph {et~al.}(2003)\citenamefont
  {Eschner}, \citenamefont {Morigi}, \citenamefont {Schmidt-Kaler},\ and\
  \citenamefont {Blatt}}]{eschner2003}%
  \BibitemOpen
  \bibfield  {author} {\bibinfo {author} {\bibfnamefont {J.}~\bibnamefont
  {Eschner}}, \bibinfo {author} {\bibfnamefont {G.}~\bibnamefont {Morigi}},
  \bibinfo {author} {\bibfnamefont {F.}~\bibnamefont {Schmidt-Kaler}},\ and\
  \bibinfo {author} {\bibfnamefont {R.}~\bibnamefont {Blatt}},\ }\href
  {https://doi.org/10.1364/JOSAB.20.001003} {\bibfield  {journal} {\bibinfo
  {journal} {J. Opt. Soc. Am. B}\ }\textbf {\bibinfo {volume} {20}},\ \bibinfo
  {pages} {1003} (\bibinfo {year} {2003})}\BibitemShut {NoStop}%
\bibitem [{\citenamefont {Delaney}\ \emph {et~al.}(2024)\citenamefont
  {Delaney}, \citenamefont {Sletten}, \citenamefont {Cich}, \citenamefont
  {Estey}, \citenamefont {Fabrikant}, \citenamefont {Hayes}, \citenamefont
  {Hoffman}, \citenamefont {Hostetter}, \citenamefont {Langer}, \citenamefont
  {Moses}, \citenamefont {Perry}, \citenamefont {Peterson}, \citenamefont
  {Schaffer}, \citenamefont {Volin}, \citenamefont {Vittorini},\ and\
  \citenamefont {Burton}}]{delaney2024}%
  \BibitemOpen
  \bibfield  {author} {\bibinfo {author} {\bibfnamefont {R.~D.}\ \bibnamefont
  {Delaney}}, \bibinfo {author} {\bibfnamefont {L.~R.}\ \bibnamefont
  {Sletten}}, \bibinfo {author} {\bibfnamefont {M.~J.}\ \bibnamefont {Cich}},
  \bibinfo {author} {\bibfnamefont {B.}~\bibnamefont {Estey}}, \bibinfo
  {author} {\bibfnamefont {M.~I.}\ \bibnamefont {Fabrikant}}, \bibinfo {author}
  {\bibfnamefont {D.}~\bibnamefont {Hayes}}, \bibinfo {author} {\bibfnamefont
  {I.~M.}\ \bibnamefont {Hoffman}}, \bibinfo {author} {\bibfnamefont
  {J.}~\bibnamefont {Hostetter}}, \bibinfo {author} {\bibfnamefont
  {C.}~\bibnamefont {Langer}}, \bibinfo {author} {\bibfnamefont {S.~A.}\
  \bibnamefont {Moses}}, \bibinfo {author} {\bibfnamefont {A.~R.}\ \bibnamefont
  {Perry}}, \bibinfo {author} {\bibfnamefont {T.~A.}\ \bibnamefont {Peterson}},
  \bibinfo {author} {\bibfnamefont {A.}~\bibnamefont {Schaffer}}, \bibinfo
  {author} {\bibfnamefont {C.}~\bibnamefont {Volin}}, \bibinfo {author}
  {\bibfnamefont {G.}~\bibnamefont {Vittorini}},\ and\ \bibinfo {author}
  {\bibfnamefont {W.~C.}\ \bibnamefont {Burton}},\ }\href
  {https://doi.org/10.1103/PhysRevX.14.041028} {\bibfield  {journal} {\bibinfo
  {journal} {Phys. Rev. X}\ }\textbf {\bibinfo {volume} {14}},\ \bibinfo
  {pages} {041028} (\bibinfo {year} {2024})}\BibitemShut {NoStop}%
\bibitem [{\citenamefont {Hankin}\ \emph {et~al.}(2019)\citenamefont {Hankin},
  \citenamefont {Clements}, \citenamefont {Huang}, \citenamefont {Brewer},
  \citenamefont {Chen}, \citenamefont {Chou}, \citenamefont {Hume},\ and\
  \citenamefont {Leibrandt}}]{hankin2019}%
  \BibitemOpen
  \bibfield  {author} {\bibinfo {author} {\bibfnamefont {A.~M.}\ \bibnamefont
  {Hankin}}, \bibinfo {author} {\bibfnamefont {E.~R.}\ \bibnamefont
  {Clements}}, \bibinfo {author} {\bibfnamefont {Y.}~\bibnamefont {Huang}},
  \bibinfo {author} {\bibfnamefont {S.~M.}\ \bibnamefont {Brewer}}, \bibinfo
  {author} {\bibfnamefont {J.-S.}\ \bibnamefont {Chen}}, \bibinfo {author}
  {\bibfnamefont {C.~W.}\ \bibnamefont {Chou}}, \bibinfo {author}
  {\bibfnamefont {D.~B.}\ \bibnamefont {Hume}},\ and\ \bibinfo {author}
  {\bibfnamefont {D.~R.}\ \bibnamefont {Leibrandt}},\ }\href
  {https://doi.org/10.1103/PhysRevA.100.033419} {\bibfield  {journal} {\bibinfo
   {journal} {Phys. Rev. A}\ }\textbf {\bibinfo {volume} {100}},\ \bibinfo
  {pages} {033419} (\bibinfo {year} {2019})}\BibitemShut {NoStop}%
\bibitem [{\citenamefont {van Mourik}\ \emph {et~al.}(2022)\citenamefont {van
  Mourik}, \citenamefont {Hrmo}, \citenamefont {Gerster}, \citenamefont
  {Wilhelm}, \citenamefont {Blatt}, \citenamefont {Schindler},\ and\
  \citenamefont {Monz}}]{Mourik2021}%
  \BibitemOpen
  \bibfield  {author} {\bibinfo {author} {\bibfnamefont {M.~W.}\ \bibnamefont
  {van Mourik}}, \bibinfo {author} {\bibfnamefont {P.}~\bibnamefont {Hrmo}},
  \bibinfo {author} {\bibfnamefont {L.}~\bibnamefont {Gerster}}, \bibinfo
  {author} {\bibfnamefont {B.}~\bibnamefont {Wilhelm}}, \bibinfo {author}
  {\bibfnamefont {R.}~\bibnamefont {Blatt}}, \bibinfo {author} {\bibfnamefont
  {P.}~\bibnamefont {Schindler}},\ and\ \bibinfo {author} {\bibfnamefont
  {T.}~\bibnamefont {Monz}},\ }\href
  {https://doi.org/10.1103/PhysRevA.105.033101} {\bibfield  {journal} {\bibinfo
   {journal} {Phys. Rev. A}\ }\textbf {\bibinfo {volume} {105}},\ \bibinfo
  {pages} {033101} (\bibinfo {year} {2022})}\BibitemShut {NoStop}%
\bibitem [{\citenamefont {Stenholm}(1986)}]{Stenholm1986}%
  \BibitemOpen
  \bibfield  {author} {\bibinfo {author} {\bibfnamefont {S.}~\bibnamefont
  {Stenholm}},\ }\href {https://doi.org/10.1103/RevModPhys.58.699} {\bibfield
  {journal} {\bibinfo  {journal} {Rev. Mod. Phys.}\ }\textbf {\bibinfo {volume}
  {58}},\ \bibinfo {pages} {699} (\bibinfo {year} {1986})}\BibitemShut
  {NoStop}%
\bibitem [{\citenamefont {Metcalf}\ and\ \citenamefont
  {Straten}(2002)}]{metcalfText}%
  \BibitemOpen
  \bibfield  {author} {\bibinfo {author} {\bibfnamefont {H.~J.}\ \bibnamefont
  {Metcalf}}\ and\ \bibinfo {author} {\bibfnamefont {P.~v.~d.}\ \bibnamefont
  {Straten}},\ }\href@noop {} {\emph {\bibinfo {title} {Laser cooling and
  trapping}}}\ (\bibinfo  {publisher} {Springer},\ \bibinfo {year}
  {2002})\BibitemShut {NoStop}%
\bibitem [{\citenamefont {Wesenberg}\ \emph {et~al.}(2007)\citenamefont
  {Wesenberg}, \citenamefont {Epstein}, \citenamefont {Leibfried},
  \citenamefont {Blakestad}, \citenamefont {Britton}, \citenamefont {Home},
  \citenamefont {Itano}, \citenamefont {Jost}, \citenamefont {Knill},
  \citenamefont {Langer} \emph {et~al.}}]{fluorescenceTheory}%
  \BibitemOpen
  \bibfield  {author} {\bibinfo {author} {\bibfnamefont {J.~H.}\ \bibnamefont
  {Wesenberg}}, \bibinfo {author} {\bibfnamefont {R.~J.}\ \bibnamefont
  {Epstein}}, \bibinfo {author} {\bibfnamefont {D.}~\bibnamefont {Leibfried}},
  \bibinfo {author} {\bibfnamefont {R.~B.}\ \bibnamefont {Blakestad}}, \bibinfo
  {author} {\bibfnamefont {J.}~\bibnamefont {Britton}}, \bibinfo {author}
  {\bibfnamefont {J.~P.}\ \bibnamefont {Home}}, \bibinfo {author}
  {\bibfnamefont {W.~M.}\ \bibnamefont {Itano}}, \bibinfo {author}
  {\bibfnamefont {J.~D.}\ \bibnamefont {Jost}}, \bibinfo {author}
  {\bibfnamefont {E.}~\bibnamefont {Knill}}, \bibinfo {author} {\bibfnamefont
  {C.}~\bibnamefont {Langer}}, \emph {et~al.},\ }\href
  {https://doi.org/10.1103/PhysRevA.76.053416} {\bibfield  {journal} {\bibinfo
  {journal} {Phys. Rev. A}\ }\textbf {\bibinfo {volume} {76}},\ \bibinfo
  {pages} {053416} (\bibinfo {year} {2007})}\BibitemShut {NoStop}%
\bibitem [{\citenamefont {Dalibard}\ and\ \citenamefont
  {Cohen-Tannoudji}(1985)}]{dalibard1985}%
  \BibitemOpen
  \bibfield  {author} {\bibinfo {author} {\bibfnamefont {J.}~\bibnamefont
  {Dalibard}}\ and\ \bibinfo {author} {\bibfnamefont {C.}~\bibnamefont
  {Cohen-Tannoudji}},\ }\href {https://doi.org/10.1088/0022-3700/18/8/019}
  {\bibfield  {journal} {\bibinfo  {journal} {Journal of Physics B: Atomic and
  Molecular Physics}\ }\textbf {\bibinfo {volume} {18}},\ \bibinfo {pages}
  {1661} (\bibinfo {year} {1985})}\BibitemShut {NoStop}%
\bibitem [{\citenamefont {Molmer}(1994)}]{molmer1994}%
  \BibitemOpen
  \bibfield  {author} {\bibinfo {author} {\bibfnamefont {K.}~\bibnamefont
  {Molmer}},\ }\href {https://doi.org/10.1088/0953-4075/27/9/024} {\bibfield
  {journal} {\bibinfo  {journal} {Journal of Physics B: Atomic, Molecular and
  Optical Physics}\ }\textbf {\bibinfo {volume} {27}},\ \bibinfo {pages} {1889}
  (\bibinfo {year} {1994})}\BibitemShut {NoStop}%
\bibitem [{\citenamefont {Morigi}\ \emph {et~al.}(1999)\citenamefont {Morigi},
  \citenamefont {Eschner}, \citenamefont {Cirac},\ and\ \citenamefont
  {Zoller}}]{morigi1999}%
  \BibitemOpen
  \bibfield  {author} {\bibinfo {author} {\bibfnamefont {G.}~\bibnamefont
  {Morigi}}, \bibinfo {author} {\bibfnamefont {J.}~\bibnamefont {Eschner}},
  \bibinfo {author} {\bibfnamefont {J.~I.}\ \bibnamefont {Cirac}},\ and\
  \bibinfo {author} {\bibfnamefont {P.}~\bibnamefont {Zoller}},\ }\href
  {https://doi.org/10.1103/PhysRevA.59.3797} {\bibfield  {journal} {\bibinfo
  {journal} {Phys. Rev. A}\ }\textbf {\bibinfo {volume} {59}},\ \bibinfo
  {pages} {3797} (\bibinfo {year} {1999})}\BibitemShut {NoStop}%
\bibitem [{\citenamefont {Morigi}\ and\ \citenamefont
  {Eschner}(2001)}]{morigi2001}%
  \BibitemOpen
  \bibfield  {author} {\bibinfo {author} {\bibfnamefont {G.}~\bibnamefont
  {Morigi}}\ and\ \bibinfo {author} {\bibfnamefont {J.}~\bibnamefont
  {Eschner}},\ }\href {https://doi.org/10.1103/PhysRevA.64.063407} {\bibfield
  {journal} {\bibinfo  {journal} {Phys. Rev. A}\ }\textbf {\bibinfo {volume}
  {64}},\ \bibinfo {pages} {063407} (\bibinfo {year} {2001})}\BibitemShut
  {NoStop}%
\bibitem [{\citenamefont {Rabl}(2010)}]{rabl2010}%
  \BibitemOpen
  \bibfield  {author} {\bibinfo {author} {\bibfnamefont {P.}~\bibnamefont
  {Rabl}},\ }\href {https://doi.org/10.1103/PhysRevB.82.165320} {\bibfield
  {journal} {\bibinfo  {journal} {Phys. Rev. B}\ }\textbf {\bibinfo {volume}
  {82}},\ \bibinfo {pages} {165320} (\bibinfo {year} {2010})}\BibitemShut
  {NoStop}%
\bibitem [{\citenamefont {Lau}\ and\ \citenamefont
  {Plenio}(2016)}]{highTempOscillator}%
  \BibitemOpen
  \bibfield  {author} {\bibinfo {author} {\bibfnamefont {H.-K.}\ \bibnamefont
  {Lau}}\ and\ \bibinfo {author} {\bibfnamefont {M.~B.}\ \bibnamefont
  {Plenio}},\ }\href {https://doi.org/10.1103/PhysRevB.94.054305} {\bibfield
  {journal} {\bibinfo  {journal} {Phys. Rev. B}\ }\textbf {\bibinfo {volume}
  {94}},\ \bibinfo {pages} {054305} (\bibinfo {year} {2016})}\BibitemShut
  {NoStop}%
\bibitem [{\citenamefont {Cirac}\ \emph {et~al.}(1992)\citenamefont {Cirac},
  \citenamefont {Blatt}, \citenamefont {Zoller},\ and\ \citenamefont
  {Phillips}}]{Cirac1992}%
  \BibitemOpen
  \bibfield  {author} {\bibinfo {author} {\bibfnamefont {J.~I.}\ \bibnamefont
  {Cirac}}, \bibinfo {author} {\bibfnamefont {R.}~\bibnamefont {Blatt}},
  \bibinfo {author} {\bibfnamefont {P.}~\bibnamefont {Zoller}},\ and\ \bibinfo
  {author} {\bibfnamefont {W.~D.}\ \bibnamefont {Phillips}},\ }\href
  {https://doi.org/10.1103/PhysRevA.46.2668} {\bibfield  {journal} {\bibinfo
  {journal} {Phys. Rev. A}\ }\textbf {\bibinfo {volume} {46}},\ \bibinfo
  {pages} {2668} (\bibinfo {year} {1992})}\BibitemShut {NoStop}%
\bibitem [{\citenamefont {Morigi}\ \emph {et~al.}(2000)\citenamefont {Morigi},
  \citenamefont {Eschner},\ and\ \citenamefont {Keitel}}]{morigi2000}%
  \BibitemOpen
  \bibfield  {author} {\bibinfo {author} {\bibfnamefont {G.}~\bibnamefont
  {Morigi}}, \bibinfo {author} {\bibfnamefont {J.}~\bibnamefont {Eschner}},\
  and\ \bibinfo {author} {\bibfnamefont {C.~H.}\ \bibnamefont {Keitel}},\
  }\href {https://doi.org/10.1103/PhysRevLett.85.4458} {\bibfield  {journal}
  {\bibinfo  {journal} {Phys. Rev. Lett.}\ }\textbf {\bibinfo {volume} {85}},\
  \bibinfo {pages} {4458} (\bibinfo {year} {2000})}\BibitemShut {NoStop}%
\bibitem [{\citenamefont {Rasmusson}\ \emph {et~al.}(2021)\citenamefont
  {Rasmusson}, \citenamefont {D'Onofrio}, \citenamefont {Xie}, \citenamefont
  {Cui},\ and\ \citenamefont {Richerme}}]{Rasmusson2021}%
  \BibitemOpen
  \bibfield  {author} {\bibinfo {author} {\bibfnamefont {A.~J.}\ \bibnamefont
  {Rasmusson}}, \bibinfo {author} {\bibfnamefont {M.}~\bibnamefont
  {D'Onofrio}}, \bibinfo {author} {\bibfnamefont {Y.}~\bibnamefont {Xie}},
  \bibinfo {author} {\bibfnamefont {J.}~\bibnamefont {Cui}},\ and\ \bibinfo
  {author} {\bibfnamefont {P.}~\bibnamefont {Richerme}},\ }\href
  {https://doi.org/10.1103/PhysRevA.104.043108} {\bibfield  {journal} {\bibinfo
   {journal} {Phys. Rev. A}\ }\textbf {\bibinfo {volume} {104}},\ \bibinfo
  {pages} {043108} (\bibinfo {year} {2021})}\BibitemShut {NoStop}%
\bibitem [{\citenamefont {Cook}(1979)}]{cook1979}%
  \BibitemOpen
  \bibfield  {author} {\bibinfo {author} {\bibfnamefont {R.~J.}\ \bibnamefont
  {Cook}},\ }\href {https://doi.org/10.1103/PhysRevA.20.224} {\bibfield
  {journal} {\bibinfo  {journal} {Phys. Rev. A}\ }\textbf {\bibinfo {volume}
  {20}},\ \bibinfo {pages} {224} (\bibinfo {year} {1979})}\BibitemShut
  {NoStop}%
\bibitem [{\citenamefont {Allevi}\ \emph {et~al.}(2013)\citenamefont {Allevi},
  \citenamefont {Bondani}, \citenamefont {Marian}, \citenamefont {Marian},\
  and\ \citenamefont {Olivares}}]{allevi2013}%
  \BibitemOpen
  \bibfield  {author} {\bibinfo {author} {\bibfnamefont {A.}~\bibnamefont
  {Allevi}}, \bibinfo {author} {\bibfnamefont {M.}~\bibnamefont {Bondani}},
  \bibinfo {author} {\bibfnamefont {P.}~\bibnamefont {Marian}}, \bibinfo
  {author} {\bibfnamefont {T.~A.}\ \bibnamefont {Marian}},\ and\ \bibinfo
  {author} {\bibfnamefont {S.}~\bibnamefont {Olivares}},\ }\href
  {https://doi.org/10.1364/JOSAB.30.002621} {\bibfield  {journal} {\bibinfo
  {journal} {J. Opt. Soc. Am. B}\ }\textbf {\bibinfo {volume} {30}},\ \bibinfo
  {pages} {2621} (\bibinfo {year} {2013})}\BibitemShut {NoStop}%
\bibitem [{\citenamefont {Wineland}\ and\ \citenamefont
  {Itano}(1979)}]{wineland1979}%
  \BibitemOpen
  \bibfield  {author} {\bibinfo {author} {\bibfnamefont {D.~J.}\ \bibnamefont
  {Wineland}}\ and\ \bibinfo {author} {\bibfnamefont {W.~M.}\ \bibnamefont
  {Itano}},\ }\href {https://doi.org/10.1103/PhysRevA.20.1521} {\bibfield
  {journal} {\bibinfo  {journal} {Phys. Rev. A}\ }\textbf {\bibinfo {volume}
  {20}},\ \bibinfo {pages} {1521} (\bibinfo {year} {1979})}\BibitemShut
  {NoStop}%
\bibitem [{\citenamefont {Hu}\ \emph {et~al.}(2011)\citenamefont {Hu},
  \citenamefont {Yang}, \citenamefont {Xu}, \citenamefont {Zhou}, \citenamefont
  {Chen}, \citenamefont {Gao}, \citenamefont {Feng},\ and\ \citenamefont
  {Lee}}]{hu2011}%
  \BibitemOpen
  \bibfield  {author} {\bibinfo {author} {\bibfnamefont {Y.~M.}\ \bibnamefont
  {Hu}}, \bibinfo {author} {\bibfnamefont {W.~L.}\ \bibnamefont {Yang}},
  \bibinfo {author} {\bibfnamefont {Y.~Y.}\ \bibnamefont {Xu}}, \bibinfo
  {author} {\bibfnamefont {F.}~\bibnamefont {Zhou}}, \bibinfo {author}
  {\bibfnamefont {L.}~\bibnamefont {Chen}}, \bibinfo {author} {\bibfnamefont
  {K.~L.}\ \bibnamefont {Gao}}, \bibinfo {author} {\bibfnamefont
  {M.}~\bibnamefont {Feng}},\ and\ \bibinfo {author} {\bibfnamefont
  {C.}~\bibnamefont {Lee}},\ }\href
  {https://doi.org/10.1088/1367-2630/13/5/053037} {\bibfield  {journal}
  {\bibinfo  {journal} {New Journal of Physics}\ }\textbf {\bibinfo {volume}
  {13}},\ \bibinfo {pages} {053037} (\bibinfo {year} {2011})}\BibitemShut
  {NoStop}%
\bibitem [{\citenamefont {Peik}\ \emph {et~al.}(1999)\citenamefont {Peik},
  \citenamefont {Abel}, \citenamefont {Becker}, \citenamefont {von Zanthier},\
  and\ \citenamefont {Walther}}]{peik1999}%
  \BibitemOpen
  \bibfield  {author} {\bibinfo {author} {\bibfnamefont {E.}~\bibnamefont
  {Peik}}, \bibinfo {author} {\bibfnamefont {J.}~\bibnamefont {Abel}}, \bibinfo
  {author} {\bibfnamefont {T.}~\bibnamefont {Becker}}, \bibinfo {author}
  {\bibfnamefont {J.}~\bibnamefont {von Zanthier}},\ and\ \bibinfo {author}
  {\bibfnamefont {H.}~\bibnamefont {Walther}},\ }\href
  {https://doi.org/10.1103/PhysRevA.60.439} {\bibfield  {journal} {\bibinfo
  {journal} {Phys. Rev. A}\ }\textbf {\bibinfo {volume} {60}},\ \bibinfo
  {pages} {439} (\bibinfo {year} {1999})}\BibitemShut {NoStop}%
\bibitem [{\citenamefont {Meekhof}\ \emph {et~al.}(1996)\citenamefont
  {Meekhof}, \citenamefont {Monroe}, \citenamefont {King}, \citenamefont
  {Itano},\ and\ \citenamefont {Wineland}}]{meekhof1996}%
  \BibitemOpen
  \bibfield  {author} {\bibinfo {author} {\bibfnamefont {D.~M.}\ \bibnamefont
  {Meekhof}}, \bibinfo {author} {\bibfnamefont {C.}~\bibnamefont {Monroe}},
  \bibinfo {author} {\bibfnamefont {B.~E.}\ \bibnamefont {King}}, \bibinfo
  {author} {\bibfnamefont {W.~M.}\ \bibnamefont {Itano}},\ and\ \bibinfo
  {author} {\bibfnamefont {D.~J.}\ \bibnamefont {Wineland}},\ }\href
  {https://doi.org/10.1103/PhysRevLett.76.1796} {\bibfield  {journal} {\bibinfo
   {journal} {Phys. Rev. Lett.}\ }\textbf {\bibinfo {volume} {76}},\ \bibinfo
  {pages} {1796} (\bibinfo {year} {1996})}\BibitemShut {NoStop}%
\bibitem [{\citenamefont {Kepesidis}\ \emph {et~al.}(2013)\citenamefont
  {Kepesidis}, \citenamefont {Bennett}, \citenamefont {Portolan}, \citenamefont
  {Lukin},\ and\ \citenamefont {Rabl}}]{kepesidis2013}%
  \BibitemOpen
  \bibfield  {author} {\bibinfo {author} {\bibfnamefont {K.~V.}\ \bibnamefont
  {Kepesidis}}, \bibinfo {author} {\bibfnamefont {S.~D.}\ \bibnamefont
  {Bennett}}, \bibinfo {author} {\bibfnamefont {S.}~\bibnamefont {Portolan}},
  \bibinfo {author} {\bibfnamefont {M.~D.}\ \bibnamefont {Lukin}},\ and\
  \bibinfo {author} {\bibfnamefont {P.}~\bibnamefont {Rabl}},\ }\href
  {https://doi.org/10.1103/PhysRevB.88.064105} {\bibfield  {journal} {\bibinfo
  {journal} {Phys. Rev. B}\ }\textbf {\bibinfo {volume} {88}},\ \bibinfo
  {pages} {064105} (\bibinfo {year} {2013})}\BibitemShut {NoStop}%
\bibitem [{\citenamefont {Carmichael}(1999)}]{carmichaelvol1}%
  \BibitemOpen
  \bibfield  {author} {\bibinfo {author} {\bibfnamefont {H.}~\bibnamefont
  {Carmichael}},\ }\href {https://books.google.com/books?id=ocgRgM-yJacC}
  {\emph {\bibinfo {title} {Statistical Methods in Quantum Optics 1: Master
  Equations and Fokker-Planck Equations}}},\ Physics and astronomy online
  library\ (\bibinfo  {publisher} {Springer},\ \bibinfo {year}
  {1999})\BibitemShut {NoStop}%
\bibitem [{\citenamefont {Home}(2006)}]{homeThesis}%
  \BibitemOpen
  \bibfield  {author} {\bibinfo {author} {\bibfnamefont {J.}~\bibnamefont
  {Home}},\ }\emph {\bibinfo {title} {Entanglement of Two Trapped-Ion Spin
  Qubits}},\ \href@noop {} {Ph.D. thesis},\ \bibinfo  {school} {University of
  Oxford} (\bibinfo {year} {2006})\BibitemShut {NoStop}%
\bibitem [{\citenamefont {Leibfried}\ \emph {et~al.}(2003)\citenamefont
  {Leibfried}, \citenamefont {Blatt}, \citenamefont {Monroe},\ and\
  \citenamefont {Wineland}}]{leibfried2003}%
  \BibitemOpen
  \bibfield  {author} {\bibinfo {author} {\bibfnamefont {D.}~\bibnamefont
  {Leibfried}}, \bibinfo {author} {\bibfnamefont {R.}~\bibnamefont {Blatt}},
  \bibinfo {author} {\bibfnamefont {C.}~\bibnamefont {Monroe}},\ and\ \bibinfo
  {author} {\bibfnamefont {D.}~\bibnamefont {Wineland}},\ }\href
  {https://doi.org/10.1103/RevModPhys.75.281} {\bibfield  {journal} {\bibinfo
  {journal} {Rev. Mod. Phys.}\ }\textbf {\bibinfo {volume} {75}},\ \bibinfo
  {pages} {281} (\bibinfo {year} {2003})}\BibitemShut {NoStop}%
\bibitem [{\citenamefont {van Enk}\ and\ \citenamefont
  {Nienhuis}(1992)}]{enk1992}%
  \BibitemOpen
  \bibfield  {author} {\bibinfo {author} {\bibfnamefont {S.~J.}\ \bibnamefont
  {van Enk}}\ and\ \bibinfo {author} {\bibfnamefont {G.}~\bibnamefont
  {Nienhuis}},\ }\href {https://doi.org/10.1103/PhysRevA.46.1438} {\bibfield
  {journal} {\bibinfo  {journal} {Phys. Rev. A}\ }\textbf {\bibinfo {volume}
  {46}},\ \bibinfo {pages} {1438} (\bibinfo {year} {1992})}\BibitemShut
  {NoStop}%
\bibitem [{\citenamefont {Bartolotta}\ \emph {et~al.}(2022)\citenamefont
  {Bartolotta}, \citenamefont {J\"ager}, \citenamefont {Reilly}, \citenamefont
  {Norcia}, \citenamefont {Thompson}, \citenamefont {Smith},\ and\
  \citenamefont {Holland}}]{bartolottaEntropy}%
  \BibitemOpen
  \bibfield  {author} {\bibinfo {author} {\bibfnamefont {J.~P.}\ \bibnamefont
  {Bartolotta}}, \bibinfo {author} {\bibfnamefont {S.~B.}\ \bibnamefont
  {J\"ager}}, \bibinfo {author} {\bibfnamefont {J.~T.}\ \bibnamefont {Reilly}},
  \bibinfo {author} {\bibfnamefont {M.~A.}\ \bibnamefont {Norcia}}, \bibinfo
  {author} {\bibfnamefont {J.~K.}\ \bibnamefont {Thompson}}, \bibinfo {author}
  {\bibfnamefont {G.}~\bibnamefont {Smith}},\ and\ \bibinfo {author}
  {\bibfnamefont {M.~J.}\ \bibnamefont {Holland}},\ }\href
  {https://doi.org/10.1103/PhysRevResearch.4.013218} {\bibfield  {journal}
  {\bibinfo  {journal} {Phys. Rev. Res.}\ }\textbf {\bibinfo {volume} {4}},\
  \bibinfo {pages} {013218} (\bibinfo {year} {2022})}\BibitemShut {NoStop}%
\bibitem [{\citenamefont {Reiter}\ and\ \citenamefont
  {S\o{}rensen}(2012)}]{reiter2012}%
  \BibitemOpen
  \bibfield  {author} {\bibinfo {author} {\bibfnamefont {F.}~\bibnamefont
  {Reiter}}\ and\ \bibinfo {author} {\bibfnamefont {A.~S.}\ \bibnamefont
  {S\o{}rensen}},\ }\href {https://doi.org/10.1103/PhysRevA.85.032111}
  {\bibfield  {journal} {\bibinfo  {journal} {Phys. Rev. A}\ }\textbf {\bibinfo
  {volume} {85}},\ \bibinfo {pages} {032111} (\bibinfo {year}
  {2012})}\BibitemShut {NoStop}%
\bibitem [{\citenamefont {Marquet}\ \emph {et~al.}(2003)\citenamefont
  {Marquet}, \citenamefont {Schmidt-Kaler},\ and\ \citenamefont
  {James}}]{crossKerr}%
  \BibitemOpen
  \bibfield  {author} {\bibinfo {author} {\bibfnamefont {C.}~\bibnamefont
  {Marquet}}, \bibinfo {author} {\bibfnamefont {F.}~\bibnamefont
  {Schmidt-Kaler}},\ and\ \bibinfo {author} {\bibfnamefont {D.~F.~V.}\
  \bibnamefont {James}},\ }\href {https://doi.org/10.1007/s00340-003-1097-7}
  {\bibfield  {journal} {\bibinfo  {journal} {Applied Physics B}\ }\textbf
  {\bibinfo {volume} {76}},\ \bibinfo {pages} {199} (\bibinfo {year}
  {2003})}\BibitemShut {NoStop}%
\bibitem [{\citenamefont {He}\ \emph {et~al.}(2017)\citenamefont {He},
  \citenamefont {Tengdin}, \citenamefont {Anderson}, \citenamefont {Rey},\ and\
  \citenamefont {Holland}}]{he2017}%
  \BibitemOpen
  \bibfield  {author} {\bibinfo {author} {\bibfnamefont {P.}~\bibnamefont
  {He}}, \bibinfo {author} {\bibfnamefont {P.~M.}\ \bibnamefont {Tengdin}},
  \bibinfo {author} {\bibfnamefont {D.~Z.}\ \bibnamefont {Anderson}}, \bibinfo
  {author} {\bibfnamefont {A.~M.}\ \bibnamefont {Rey}},\ and\ \bibinfo {author}
  {\bibfnamefont {M.}~\bibnamefont {Holland}},\ }\href
  {https://doi.org/10.1103/PhysRevA.95.053403} {\bibfield  {journal} {\bibinfo
  {journal} {Phys. Rev. A}\ }\textbf {\bibinfo {volume} {95}},\ \bibinfo
  {pages} {053403} (\bibinfo {year} {2017})}\BibitemShut {NoStop}%
\bibitem [{\citenamefont {M{\o}lmer}\ \emph {et~al.}(1993)\citenamefont
  {M{\o}lmer}, \citenamefont {Castin},\ and\ \citenamefont
  {Dalibard}}]{molmer1993}%
  \BibitemOpen
  \bibfield  {author} {\bibinfo {author} {\bibfnamefont {K.}~\bibnamefont
  {M{\o}lmer}}, \bibinfo {author} {\bibfnamefont {Y.}~\bibnamefont {Castin}},\
  and\ \bibinfo {author} {\bibfnamefont {J.}~\bibnamefont {Dalibard}},\ }\href
  {https://doi.org/10.1364/JOSAB.10.000524} {\bibfield  {journal} {\bibinfo
  {journal} {J. Opt. Soc. Am. B}\ }\textbf {\bibinfo {volume} {10}},\ \bibinfo
  {pages} {524} (\bibinfo {year} {1993})}\BibitemShut {NoStop}%
\bibitem [{\citenamefont {Surajit~Sen}\ and\ \citenamefont
  {Gangopadhyay}(2015)}]{sen2015}%
  \BibitemOpen
  \bibfield  {author} {\bibinfo {author} {\bibfnamefont {M.~R.~N.}\
  \bibnamefont {Surajit~Sen}, \bibfnamefont {Tushar Kanti~Dey}}\ and\ \bibinfo
  {author} {\bibfnamefont {G.}~\bibnamefont {Gangopadhyay}},\ }\href
  {https://doi.org/10.1080/09500340.2014.960019} {\bibfield  {journal}
  {\bibinfo  {journal} {Journal of Modern Optics}\ }\textbf {\bibinfo {volume}
  {62}},\ \bibinfo {pages} {166} (\bibinfo {year} {2015})},\ \Eprint
  {https://arxiv.org/abs/https://doi.org/10.1080/09500340.2014.960019}
  {https://doi.org/10.1080/09500340.2014.960019} \BibitemShut {NoStop}%
\bibitem [{\citenamefont {Morigi}(2003)}]{morigi2003}%
  \BibitemOpen
  \bibfield  {author} {\bibinfo {author} {\bibfnamefont {G.}~\bibnamefont
  {Morigi}},\ }\href {https://doi.org/10.1103/PhysRevA.67.033402} {\bibfield
  {journal} {\bibinfo  {journal} {Phys. Rev. A}\ }\textbf {\bibinfo {volume}
  {67}},\ \bibinfo {pages} {033402} (\bibinfo {year} {2003})}\BibitemShut
  {NoStop}%
\bibitem [{\citenamefont {J.~Oz-Vogt}\ and\ \citenamefont
  {Revzen}(1991)}]{tdc}%
  \BibitemOpen
  \bibfield  {author} {\bibinfo {author} {\bibfnamefont {A.~M.}\ \bibnamefont
  {J.~Oz-Vogt}}\ and\ \bibinfo {author} {\bibfnamefont {M.}~\bibnamefont
  {Revzen}},\ }\href {https://doi.org/10.1080/09500349114552501} {\bibfield
  {journal} {\bibinfo  {journal} {Journal of Modern Optics}\ }\textbf {\bibinfo
  {volume} {38}},\ \bibinfo {pages} {2339} (\bibinfo {year} {1991})},\ \Eprint
  {https://arxiv.org/abs/https://doi.org/10.1080/09500349114552501}
  {https://doi.org/10.1080/09500349114552501} \BibitemShut {NoStop}%
\end{thebibliography}

\appendix

\section{Derivation of the semiclassical model} 
\label{modelDerivation}

Here we derive the semiclassical model in full detail.
We begin with a quantum master equation that describes the internal and motional dynamics of an arbitrary number of ions
\begin{equation}
\label{ME}
	\frac{ d \hat \rho}{dt} = \mathcal{L}^0 \hat \rho + \mathcal{D}\hat \rho + \mathcal{L}_\text{int}\hat \rho.
\end{equation}
The first term
\begin{equation}
\label{bareEnergy}
	\mathcal{L}^0 \hat \rho = -i \sum_j \left[ \hat H^0_{j} + \hat H^\text{mech}_j, \, \hat \rho \right]
\end{equation}
defines the energy of the system in the absence of the laser interaction, including the internal level energies
\begin{equation}
	\hat H^0_{j} = \sum_\alpha \omega_\alpha \ket{\alpha} \bra{\alpha}
\end{equation} 
and total mechanical energy 
\begin{equation}
	\hat H^\text{mech}_j \equiv \frac{\bo{p}_{j}^2}{2m_j} + V_j(\overrightharpoonup{\bo{r}},t)
\end{equation}
in the potential
\begin{align}
\label{trapPotential}
\begin{split}
	 V_j(\overrightharpoonup{ \bo r},t) 
	 & = V_\text{trap}(\bo{r}_j,t) + V_\text{Coul}(\overrightharpoonup{\bo r },t) \\
	 & = V_\text{trap}(\bo{r}_j,t) + \frac{e^2}{4\pi \epsilon_0} \sum_{i\neq j} \frac{1}{|\bo{r}_i - \bo{r}_j|}.
\end{split}
\end{align}
We leave the exact form of the trap potential $V_\text{trap}$ to be determined by the specific calculation; one can model, e.g., a pseudopotential, a time-dependent RF trap, and trap anharmonicities.
The operators $\bo{r}_j$ and $\bo{p}_j$  are the position and momentum operators for the center of mass of ion $j$ with mass $m_j$. 
The variable $\overrightharpoonup{\bo{r}} = \bo{r}_1, \bo{r}_2, \ldots $ denotates the dependence on all ion positions in the Coulomb interaction $V_\text{Coul}$.
The second term 
\begin{equation}
\mathcal{D} \hat \rho = \sum_j \mathcal{D}_j \hat \rho
\end{equation}
describes spontaneous emission of each ion $j$ into free space and any associated recoil. The last term
\begin{equation}
\label{laserInteraction}
	\mathcal{L}_\text{int} \hat \rho = -i \left[ \sum_{j,l,\alpha,\beta} \frac{\Omega_{\alpha \beta}^{(l,j)}}{2} \hat{\sigma}_{\alpha \beta}^{(j)} e^{i \bs k_l \cdot \bo{r}_j - \omega_l t} + \text{h.c.} , \, \hat \rho \right]
\end{equation}
describes the interaction between each ion $j$ and each classical (traveling wave) laser field $l$ in the rotating wave approximation. Here, $\hat{\sigma}_{\alpha\beta}^{(j)} \equiv \ket{\alpha}^{(j)}\bra{\beta}$ is a transition operator that acts on the internal Hilbert space of ion $j$, and the Rabi frequency $\Omega_{\alpha \beta}^{(l,j)}$ implicitly accounts for the details of the transition (e.g., if it is an electric dipole interaction).
For the cases we consider, we assume that the Rabi frequencies are constant in space and time, although it is straightforward to include spatio-temporal dependence.

We now calculate the expectation values
\begin{widetext}
\begin{align}
\label{ehrenfestR}
	\frac{d \braket{\bo{r}_j}}{dt} &
	=  \frac{\braket{\bo{p}_j}}{m_j}, \\
\label{ehrenfestP}
	\frac{d \braket{\bo{p}_j}}{dt} &= \sum_{j'} \braket{- \nabla_{\bo{r}_j} V_{j'}(\overrightharpoonup{\bo{r}},t)} 
		+\left(
			-i\sum_{l,\alpha,\beta} \bs k_l \frac{\Omega_{\alpha \beta}^{(l,j)}}{2} \braket{  \hat{\sigma}_{\alpha \beta}^{(j)} e^{i \bs k_l \cdot \bo{r}_j} } e^{ -i \omega_l t} 
			+ \text{c.c.} 
		\right),
\end{align}
\end{widetext}
in which we have used the relations $\braket{\hat{\mathcal O}} = \text{Tr}\left[ \hat{\mathcal O} \hat \rho \right]$ and $\bo{p}_j  = - i \nabla_{\bo{r}_j}$.
Equations~\eqref{ehrenfestR} and~\eqref{ehrenfestP} are an instance of Ehrenfest's theorem applied to a quantum master equation.
Because spontaneous emission into free space yields a zero average force, a term involving the dissipator $\tr{\mathcal{D}_{IE} \hat \rho \, \bo{p}_j}$ in Eq.~\eqref{ehrenfestP} has vanished.
If the diffusive effects of spontaneous recoil are of interest, one must consider other methods, such as a theory involving higher-order moments of $\hat{\bs x}$ and $\hat{\bs p}$~\cite{dalibard1985}, stochastic evolution~\cite{he2017,molmer1993}, or a more quantum treatment.
Here we focus on the other terms in Eq.~\eqref{ehrenfestP}, which includes average forces from the potential $V$ and the laser interactions. 

We now make two approximations, which when employed together can be considered the ``semiclassical approximation."
The first is to factorize the operator expectation values which couple the composite internal+motional Hilbert space~\cite{rabl2010}, e.g.: 
\begin{equation}
\label{semiclassical1}
	\braket{  \hat{\sigma}_{\alpha \beta}^{(j)} e^{i \bs k_l \cdot \bo{r}_j} }
	\approx 
	\braket{ \hat{\sigma}_{\alpha \beta}^{(j)} }
	\braket{e^{i \bs k_l \cdot \bo{r}_j}  },
\end{equation}
and the second is to replace all instances of the motional operators with their expectation values $\bs r_j(t) \equiv \braket{\bo{r}_j}(t)$, e.g.:
\begin{align}
\label{semiclassical2}
\begin{split}
	\braket{ e^{i \bs k_l \cdot \bo{r}_j} }
	&\approx e^{i \bs k_l \cdot \bs r_j}; \\
	\braket{\nabla_{\bo{r}_j} V_{j'}(\overrightharpoonup{\bo{r}},t)}
	&\approx \nabla_{\bs r_j} V_{j'}(\overrightharpoonup{ \bs r},t).
\end{split}
\end{align}
This mapping in the dissipator $\mathcal{D} \hat \rho$ completely removes the motional dependence, leaving only the effects of spontaneous emission on the internal subspace $\mathcal{H}^I$.

This treatment leaves us with a quantum master equation that spans only $\mathcal{H}^I$.
Next, we ignore any quantum coherences between the internal subspaces of the ions due to the laser interactions and spontaneous emission by treating the composite internal subspace as a direct sum:
\begin{equation}
	\mathcal{H} ^I = \mathcal{H}^I_1 \bigoplus \mathcal{H}^I_2 \bigoplus \ldots.
\end{equation}
Importantly, the Coulomb interaction can still cause inter-ion correlations to develop.
This simplification allows us to evolve a system of $j$ quantum master equations, coupled to one another only through the motional effects.

Combining these results, we arrive at Newton's equations of motion for the motional dynamics and quantum master equations for the internal dynamics.
Moving to the interaction picture defined by the energies of the internal levels $H^0 = \sum_j \hat H_j^0$, we find the expressions
\begin{align}
\label{Newton_A}
\begin{split}
	\frac{d \bs{r}_j}{dt} & = \frac{\bs{p}_j}{m_j}; \\
	\frac{d \bs{p}_j}{dt} & = \bs F_j^\text{laser} + \bs F_j^\text{trap} + \bs F_j^\text{Coloumb},
\end{split}
\end{align}
and
\begin{equation}
\label{internalME_A}
	\frac{d \hat \rho^I_j}{dt} =  -i \left[ \hat H_j(\bs r_j, t) , \, \hat \rho^I_j \right]
	+ \mathcal{D}_j\hat \rho^I_j.
\end{equation}
Here,
\begin{equation}
	\hat H_j(\bs r_j, t) = \sum_{l,\alpha,\beta} \frac{\Omega_{\alpha \beta}^{(l,j)}}{2} \hat{\sigma}_{\alpha \beta}^{(j)} e^{- i \phi^{(l,j)}_{\alpha \beta}(\bs r_j,t)} + \text{h.c.}
\end{equation}
is the ion-laser interaction Hamiltnoian, $\hat \rho^I_j$ is the reduced density matrix on the internal subspace of the $j^\text{th}$ ion, $\mathcal{D}_j\hat \rho^I_j$ accounts for dissipative dynamics on the internal subspace, and the instantaneous laser phase (set to zero at $t=0$) is
\begin{equation}
\label{motionalPhase}
\begin{aligned}
	\phi^{(l,j)}_{\alpha \beta}(\bs r_j, t) &=  - \bs k_l \cdot \bs r_j(t) + \int_{0}^t  \left[\omega_l - \left(\omega_\alpha - \omega_\beta \right) \right]  dt' \\
	& = \int_{0}^t  \left[\Delta_{\alpha \beta}^{(l)} - \bs k_l \cdot \bs v_j(t')\right]  dt'.
\end{aligned}
\end{equation}
We emphasize that the laser phases depend on the instantaneous laser detuning 
\begin{equation}
	\Delta_{\alpha \beta}^{(l)}(t) = \omega_l(t) - \left(\omega_\alpha - \omega_\beta \right) 
\end{equation}
(which we now allow to be time-dependent) and Doppler shift  $\bs k_l \cdot \bs v_j(t)$ witnessed by each ion.
Lastly, the forces are
\begin{align}
\label{forces_A}
	\bs F^\text{trap}_j & = - \nabla_{\bs r_j} V_\text{trap}(\bs r_j,t); \notag \\
	\bs F^\text{Coulomb}_j & = - \nabla_{\bs r_j} V_\text{Coulomb}(\overrightharpoonup{\bs r },t);  \\
	\bs F^\text{laser}_j & = -i
			\sum_{l,\alpha,\beta} \bs k_l \frac{\Omega_{\alpha \beta}^{(l,j)}}{2} \braket{  \hat{\sigma}_{\alpha \beta}^{(j)} }e^{- i  \phi^{(l,j)}_{\alpha \beta}}
			+ \text{c.c.}. \notag
\end{align}

\section{Analytic study of EIT cooling beyond the Lamb-Dicke regime}
\label{captureRange}

In Section~\ref{expComparison}, we presented numerical and experimental evidence of an EIT capture range, such that systems with sufficiently high motional excitation are heated instead of cooled.
Here we develop a simple semiclassical model to derive an analytic approximation of the capture range and the cooling rate beyond the Lamb-Dicke regime for a single motional mode.

We again consider a $\Lambda$-system, as depicted in Fig.~\ref{fig:simpleEIT}(b).
Here, we denote the two ground states as $\ket{1}, \ket{2}$, and the excited state as $\ket{e}$.
In a time-independent rotating frame, the Hamiltonian reads
\begin{equation}
	\hat{H} = \sum_{j=1,2} \Delta_j \ket{j}\bra{j} + \left( \frac{\Omega_j}{2} \ket{e}\bra{j} + \text{h.c.} \right),
\end{equation}
wherein $\Delta_j$ is the detuning of laser $j$, which couples states $\ket{j}$ and $\ket{e}$ with Rabi rate $\Omega_j = \Omega_j(\hat{\bs r})$.
The dissipative effects are described by the jump operators
\begin{equation}
	\hat{L}_j = \sqrt{\Gamma_j}\ket{j}\bra{e},
\end{equation} 
in which $\Gamma_j$ is the decay rate from $\ket{e}$ to $\ket{j}$, and $\Gamma = \Gamma_1 + \Gamma_2$ is the excited state linewidth.
As in Eq.~\eqref{motionalPhase}, we introduce classical motion by incorporating Doppler shifts as
\begin{equation}
	\Delta_j \rightarrow \Delta_j - \bs k_j \cdot \bs v(t).
\end{equation}

\subsection{Adiabatic elimination}
\label{AE}

To reduce the complexity of the resulting solutions, we adiabatically eliminate the excited state by use of an effective operator formalism~\cite{reiter2012}.
This procedure results in the effective Hamiltonian
\begin{equation}
\label{Heff}
\begin{aligned}
	\hat H_\text{eff} & = \sum_{j=1,2} \Delta_j \left( 1 + \frac{ |\Omega_j|^2}{4 \Delta_j^2  + \Gamma^2} \right) \ket{j}\bra{j}   \\
	& + \left( \frac{\Omega^\text{R}}{2} \ket{1}\bra{2} + \text{H.c.} \right)
\end{aligned}
\end{equation}
and effective jump operators
\begin{equation}
\label{Leff}
	\hat{L}_{j}^\text{eff}  = 
		- \frac{\sqrt{\Gamma_j}}{2} \ket{j}
		\sum_{l=1,2}
		\frac{\Omega_l}{\Delta_l + i \Gamma/2} \bra{l},
\end{equation}
in which
\begin{equation}
	\Omega^\text{R} =
	 \frac{(\Omega_1)^* \Omega_2 \left(	\Delta_1 + \Delta_2 \right)}
	 {4 \left(\Delta_2 + i \Gamma/2	\right) \left(\Delta_1 -  i \Gamma/2 \right)}
\end{equation}
is the Raman Rabi frequency.
The ground-state subspace then evolves according to the effective quantum master equation
\begin{equation}
\label{QMEeff}
	\frac{d \hat \rho_\text{eff}}{dt} = -i \left[ \hat H_\text{eff}, \hat \rho_\text{eff} \right] 
	+ \mathcal{D}_\text{eff} \hat \rho_\text{eff}
\end{equation}
wherein
\begin{equation}
\begin{aligned}
	\mathcal{D}_\text{eff} \hat \rho_\text{eff} & =  
	\sum_{j=1,2}
		\hat L^\text{eff}_j \hat \rho_\text{eff} \left( L^\text{eff}_j  \right)^\dag \\
		& \qquad - \frac{1}{2} \left\{ \left( L^\text{eff}_j  \right)^\dag L^\text{eff}_j, \hat \rho_\text{eff}\right\}.
\end{aligned}
\end{equation}

Following the same procedure as in Appendix~\ref{modelDerivation}, the effective laser forces are calculated to be
\begin{equation}
\label{Feff}
	\boldsymbol F^\text{eff}_\text{laser}
	=  \bs F_\text{coh} + \bs F_\text{diss},
\end{equation}
where
\begin{equation}
	\bs F_\text{coh} = \left< - \nabla \hat H_\text{eff} \right> =
		(\bs k_1 - \bs k_2)
		\left(	i \frac{\Omega^\text{R}}{2} \rho_{21} + \text{h.c.}
		\right),
\end{equation}
\begin{equation}
	\bs F_\text{diss}
	 = \text{Tr} \left( \mathcal{D}_\text{eff} \rho \, \hat{ \boldsymbol p} \right)  =
		\Gamma \sum_{j=1,2} \frac{|\Omega_j|^2}{4\Delta_j ^2+\Gamma^2}	\rho_{jj} \boldsymbol{k}_j,
\end{equation} 
and $\rho_{ij} = \braket{i|\hat \rho|j}$.
Physically, $\bs F_\text{coh}$ captures net forces due to coherent Raman transitions, which are essential in EIT and sideband cooling on a Raman transition, whereas $\bs F_\text{diss}$ captures net forces due to spontaneous Raman transitions, which are essential in Doppler cooling and the culprit for high-energy EIT heating.
Because we are considering a single mode, we can simplify Eq.~\eqref{dndt} to
\begin{equation}
\label{dndtSimp}
	R(\mathscr n) = 2 \pi \int_{- \pi/\omega}^{\pi/\omega} \bs{F}_\text{laser} (\mathscr n,t) \cdot \bs{v} (\mathscr n,t) \, dt,
\end{equation}
so that
\begin{equation}
\label{dndteff}
	R_\text{eff}(\mathscr n)  = R_\text{coh}(\mathscr n)  + R_\text{diss}(\mathscr n).
\end{equation}

The adiabatic elimination results in a computational savings and an intuitive physical picture.
To derive an analytic approximation to the cooling dynamics, we also need to consider a dynamic steady-state solution.

\subsection{Dynamic steady state}

The dynamics of an open quantum system can often be written in terms of optical Bloch equations (OBEs)~\cite{sen2015}
\begin{equation}
\label{OBEs}
	\frac{d}{dt}{\bs{S}}(t) = \bs{\mathsf{A}}(t) \bs{S}(t).
\end{equation}
In our case, $\bs S(t) = (\rho_{11}, \text{Re}[\rho_{12}], \text{Im}[\rho_{12}])^T$, and $\bs{\mathsf{A}}(t)$ is a matrix which couples the elements of $\bs S(t)$.
If the motion is harmonic at frequency $\omega$, then the solution to the OBEs is the dynamic steady state~\cite{rabl2010,highTempOscillator}
\begin{equation}
	\bs{S}(t) = 
		\lim_{q_\text{max} \rightarrow \infty} 
		\sum_{q=-q_\text{max}}^{q_\text{max}} \bs{S}^{(q)} e^{i q \omega t},
\end{equation}
in which $\bs{S}^{(q)}$ is a vector of time-independent, complex numbers.
Approximations to $\bs{S}(t)$ can be calculated by choosing a finite $q_\text{max}$, substituting into the OBEs, and solving the resulting system of time-independent algebraic equations
\begin{equation}
\label{algebraic}
	i \omega \bs{S}^{(q)} = \sum_{q'=-q_\text{max}}^{q_\text{max}} \bs{\mathsf{A}}^{(q-q')} \bs{S}^{(q')} 
\end{equation}
 for $\bs S^{(-q_\text{max})},\bs S^{(-q_\text{max}+1)}, \ldots, \bs S^{(q_\text{max})}$.
 Here, we have decomposed the coupling matrix in a similar way as
 \begin{equation}
 	\bs{\mathsf{A}}(t) = 
		\sum_{q=-\infty}^\infty \bs{\mathsf{A}}^{(q)} e^{i q \omega t},
 \end{equation}
 where $\bs{\mathsf{A}}^{(q)}$ is a time-independent matrix.
This approximate solution can then be used to calculate observables, such as laser forces and cooling rates.

\begin{figure*}
	\input{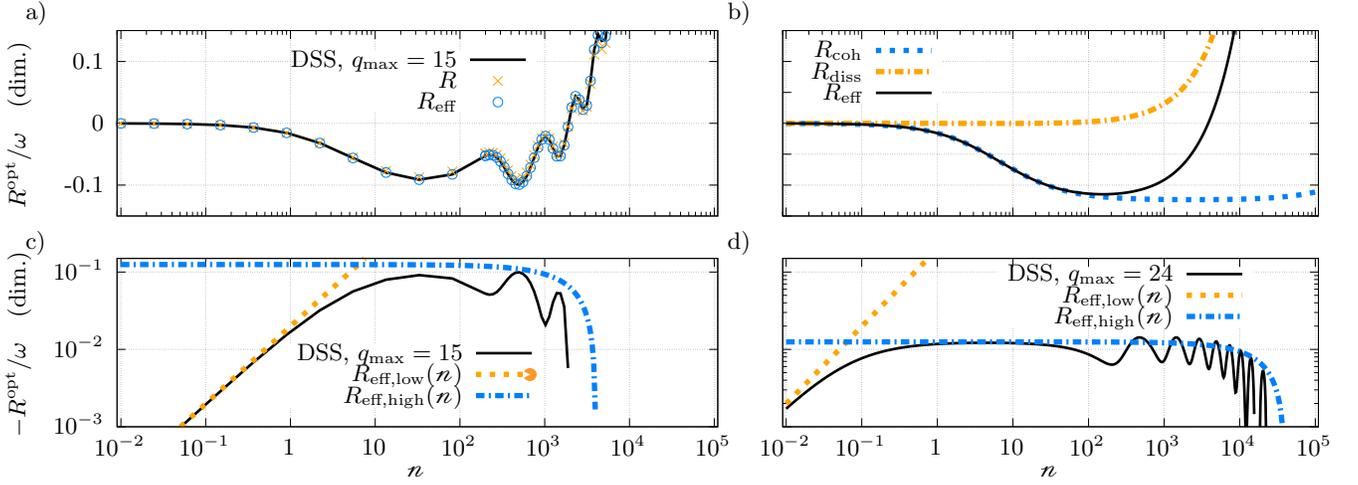}
	\caption{EIT cooling performance as a function of energy. Common parameters are: $\eta = 0.1$ and $\omega = \Gamma/20$. We set $\Delta = 2 \Gamma$ for plots a), b), and c), and $\Delta = 20 \Gamma$ for plot d). We have used the Rabi rate $\Omega_\text{opt}(\Delta)$ [Eq.~\eqref{Omegaopt}]. a) Comparing the numerical predictions from the dynamic steady-state (DSS) approach (black, solid), the PACMAN method presented in the main text [Eq.~\eqref{dndt}, orange crosses], and PACMAN including adiabatic elimination [Eq.~\eqref{dndteff}, blue circles]. b) Expressing the dynamics as the sum of a cooling term $R_\text{coh}$ (blue, dashed) and a heating term $R_\text{diss}$ (orange, dot-dashed), as shown in Eq.~\eqref{dndteff}. The approximate solution shown in Eq.~\eqref{dndteffApprox} has been used. c) EIT cooling when $\Delta = 2\Gamma$. A cutoff $q_\text{max}=15$ was necessary to reach convergence. d) Same as plot c), but with $\Delta = 20 \Gamma$. A cutoff $q_\text{max}=24$ was necessary to reach convergence.}
	\label{fig:DSS}
 \end{figure*}

In Fig.~\ref{fig:DSS}(a), we present an example that demonstrates the strong agreement among the various approaches.
That is, we numerically calculated the internal dynamics in three separate ways: simulating the quantum master equation [Eq.~\eqref{internalME}, orange crosses], simulating the effective quantum master equation [Eq.~\eqref{QMEeff}, blue circles], and solving the algebraic system  [Eq.~\eqref{algebraic}, black, solid],
which were then used to calculate $R$ and $R_\text{eff}$. 
We found the cutoff $q_\text{max}=15$ was necessary to reach convergence at high energies for the dynamic steady-state method.
Based on this result, we assume that the adiabatic elimination and dynamic steady-state ansatz sufficiently capture the EIT cooling dynamics and use their predictions in what follows.

\subsection{Analytic approximation to the cooling dynamics}

Before attempting to solve Eq.~\eqref{dndteff} analytically, we make a few simplifications.
First, we symmetrize the parameters so that $\Delta_1 = \Delta_2 = \Delta$ (operating near the EIT dark resonance), $\Omega_1 = \Omega_2 = \Omega$ (maximizing the cooling rate~\cite{morigi2003}),  $\bs k_1 = - \bs k_2 = \bs k$ (reducing the cooling to one spatial dimension), and $\Gamma_1 = \Gamma_2 = \Gamma/2$.
In particular, this allows us to express the Doppler shift as
\begin{equation}
	\bs k \cdot \bs v = \eta \omega \sqrt{\mathscr n} \cos(\omega t),
\end{equation}
in which $\eta = 2 |\bs k| $ is the two-photon Lamb-Dicke parameter, and $\mathscr n$ is the average occupancy of a coherent state with frequency $\omega$ and velocity expectation value $\bs{v}(t)$.
Next, we assume that $\Delta \gg \Gamma/2, \eta \omega \sqrt{\mathscr n}$, which allows for a Taylor expansion of $A(t)$ and $\bs F_\text{eff}$ in powers of $1/\Delta$.
This expansion is motivated by the form of the function 
\begin{equation}
\label{mFunction}
	g(t) = \frac{1}{\Gamma^2 + 4[\Delta  - \bs{k} \cdot \bs{v}(t)]^2},
\end{equation}
which parameterizes the effective coupling matrix: $\bs{\mathsf{A}}(t) = \bs{\mathsf{A}}[g(t)].$

We are now equipped to calculate an analytic approximation to the EIT cooling dynamics. 
Because the Lamb-Dicke regime cooling rate $\mathcal W$ scales like $1/\Delta^3$~\cite{morigi2000}, we perform the Taylor expansions in $1/\Delta$ to the same order.
Further, to calculate the lowest-order motional corrections to $\bs{S}(t)$, we choose $q_\text{max} = 1$.
We then solve Eq.~\eqref{algebraic} and substitute the results into Eq.~\eqref{dndteff} to find
\begin{equation}
\label{dndteffApprox}
	R_\text{eff}(\mathscr n)  \approx 
	\frac{ \sum_{k=1}^3 N_k \mathscr n^k}{\sum_{k=0}^4 D_k \mathscr n^k},
\end{equation}
in which
\begin{widetext}
\begin{align}
\label{bigEQ}
	N_1 &= 32 \Gamma  \Delta ^3 \eta ^2 \omega ^2 \Omega ^2 \left[64 \Delta ^6 \omega ^2+8 \Delta ^4 \left(\Gamma ^2-4 \Delta ^2\right) \Omega
   ^2-\left(\Gamma ^4-10 \Gamma ^2 \Delta ^2+8 \Delta ^4\right) \Omega ^4\right]; \notag \\
   	N_2 &= 64 \Gamma  \Delta ^3 \eta ^4 \omega ^4 \Omega ^2 \left[32 \Delta ^6-4 \Delta ^2 \left(\Gamma ^2+2 \Delta ^2\right) \Omega ^2+\left(3
   \Gamma ^2-8 \Delta ^2\right) \Omega ^4\right]; \notag \\
	N_3 &= 32 \Gamma  \Delta ^3 \eta ^6 \omega ^6 \Omega ^4 \left(8 \Delta ^2-5 \Omega ^2\right); \notag \\
	D_0 &= 4096 \Delta ^{12} \omega ^4-128 \Delta ^6 \left(\Gamma ^4-12 \Gamma ^2 \Delta ^2+16 \Delta ^4\right) \omega ^2 \Omega ^4+\left(\Gamma
   ^4-4 \Gamma ^2 \Delta ^2+16 \Delta ^4\right)^2 \Omega ^8;  \\
	D_1 &= 8 \Delta ^2 \eta ^2 \omega ^2 \big(1024 \Delta ^{10} \omega ^2+5 \Gamma ^4 \Omega ^8+256 \Delta ^8 \Omega ^2 \left[\Omega
   ^2-2 \omega ^2\right]-64 \Delta ^6 \Omega ^4 \left[\Gamma ^2+3 \omega ^2+2 \Omega ^2\right] \notag \\
     & \quad -4 \Delta ^2 \left[2 \Gamma ^4 \Omega ^6+5 \Gamma ^2
   \Omega ^8\right]+16 \Delta ^4 \Omega ^4 \left[\Gamma ^4+5 \Omega ^4+2 \Gamma ^2 \left(4 \omega ^2+\Omega ^2\right)\right]\big); \notag \\
	D_2 &= 2 \eta ^4 \omega ^4 \Omega ^4 \left[64 \Delta ^6 \left(8 \Delta ^2-2 \Gamma ^2-3 \omega ^2\right)+64 \Delta ^4 \left(\Gamma ^2-4 \Delta
   ^2\right) \Omega ^2+\left(3 \Gamma ^4-44 \Gamma ^2 \Delta ^2+208 \Delta ^4\right) \Omega ^4\right]; \notag \\
	D_3 &= 8 \eta ^6 \omega ^6 \Omega ^4 \left[16 \Delta ^6-8 \Delta ^4 \Omega ^2+\left(13 \Delta ^2-2 \Gamma ^2\right) \Omega ^4\right]; \notag \\
	D_4 &= 9 \eta ^8 \omega ^8 \Omega ^8. \notag 
\end{align}
\end{widetext}
Although we don't show the expressions here, this result is initially separated into coherent $R_\text{coh}$ and dissipative $R_\text{diss}$ terms, as in Eq.~\eqref{dndteff}.
We plot these terms in Fig.~\ref{fig:DSS}(b), in which we observe the transition from $|R_\text{diss}| \ll |R_\text{coh}|$ at low energies $\mathscr n < 1000$, where cooling dominates, to $|R_\text{diss}| \gg |R_\text{coh}|$ at high energies $\mathscr n > 1000$, where heating dominates.

While the cumbersome form of Eq.~\eqref{bigEQ} is not particularly illuminating, it can be simplified in certain relevant limits to estimate $R_\text{eff}$ at various energy scales.
Starting with the EIT capture range, defined as the solution to $R(\mathscr n_\text{cap}) = 0$ with $\mathscr n_\text{cap} > 0$, we find in the large $\Delta$ limit
\begin{equation}
	\mathscr n_\text{cap} \approx \frac{\Omega^2 - \omega^2}{2 \eta^2 \omega^2} + \mathcal{O}\left( \frac{1}{\Delta^2} \right).
\end{equation}
In particular, this result suggests that the EIT capture range scales with $\Omega^2$.
However, increasing only the Rabi rate can reduce the cooling rate and increase the steady-state average occupancy $\bar n_\text{ss}$.
A choice which minimizes the latter to $\bar n_\text{ss,min} = (\Gamma/4\Delta)^2$, maximizes $R$ at low energies, and fixes the Raman Rabi frequency is
\begin{equation}
\label{Omegaopt}
	\Omega_\text{opt}(\Delta) = \sqrt{2\omega(\omega+\Delta)},
\end{equation}
for which the first-order red sideband is driven resonantly~\cite{morigi2003}.
(We refer to this as the ``optimal" Rabi rate, as it brings the system nearest to its motional ground state, although one could alternatively optimize the cooling at intermediate energy.)
With this choice of Rabi rate, the capture range is
\begin{equation}
\label{ncapopt}
	\mathscr n_\text{cap}^\text{opt} \approx \frac{\Delta+\omega}{\eta^2 \omega} + \mathcal{O}\left( \frac{1}{\Delta}\right),
\end{equation}
which predicts an arbitrarily large capture range in the large $\Delta$ limit.

Next, we calculate simple approximations to $R_\text{eff}$ in the zero-energy limit and near the capture range, again using the Rabi rate $\Omega_\text{opt}$.
Expanding Eq.~\eqref{dndteffApprox} in the small $\mathscr n$ and large $\Delta$ limit,
\begin{equation}
\label{Rlown}
	R_\text{eff, low}^\text{opt}(\mathscr n) \approx - \frac{\eta^2 \omega \Gamma \Delta }{\Gamma^2 + 4\omega^2} \, \mathscr n + \mathcal{O}\left( 1\right),
\end{equation}
which is similar to the LD model prediction~\cite{morigi2003}.
Expanding Eq.~\eqref{dndteffApprox} about $\mathscr n = \mathscr n_\text{cap}$ in the large $\Delta$ limit,
\begin{equation}
\label{Rhighn}
	R_\text{eff, high}^\text{opt}(\mathscr n) \approx \frac{\eta^2 \omega^2 \Gamma }{4\Delta^2} (\mathscr n - \mathscr n_\text{cap}^\text{opt}).
\end{equation}
We compare these low-energy [Eq.~\eqref{Rlown}, orange, dashed] and high-energy [Eq.~\eqref{Rhighn}, blue, dot-dashed] predictions to the dynamic steady-state solution with a large cutoff $q_\text{max}$ (DSS, black, solid) in Figs.~\ref{fig:DSS}(c-d).
To demonstrate the strategy of increasing $\mathscr n_\text{cap}$ by increasing $\Delta$, we set $\Delta = 2 \Gamma$ in Fig.~\ref{fig:DSS}(c), and $\Delta = 20 \Gamma$ in Fig.~\ref{fig:DSS}(d).

We find $R_\text{eff, low}^\text{opt}(\mathscr n)$ to strongly agree with the DSS at sufficiently low energy.
Importantly, we observe sub-exponential cooling at lower energies as $\Delta$ increases.
We find the approximation to $\mathscr n_\text{cap}^\text{opt}$ in Eq.~\eqref{ncapopt} to overestimate the capture range by roughly a factor of two, which causes disagreement between $R_\text{eff, high}^\text{opt}(\mathscr n)$  and the DSS near the capture range.
Nevertheless, $R_\text{eff, high}^\text{opt}(\mathscr n)$ qualitatively captures the intermediate-energy cooling performance and becomes more quantitatively accurate with increasing $\Delta$.
For intermediate energies $1 \ll \mathscr n \ll n_\text{cap}$, we observe a flattening of the cooling dynamics such that 
\begin{equation}
	 R_\text{eff}^\text{opt}( 1 \ll \mathscr n \ll n_\text{cap}) 
	 \approx  R_\text{eff, high}^\text{opt}(0) 
	 \approx - \frac{\omega \Gamma}{4 \Delta},
\end{equation}
demonstrating the worsening overall cooling performance with increasing $\Delta$.
The most appropriate detuning $\Delta$ for a given application can then be chosen by considering the tradeoff among $\bar n_\text{ss} \propto 1/\Delta^2$, the intermediate cooling performance $\propto 1/\Delta$, the capture range $\mathscr n_\text{cap} \propto \Delta$, and the available laser power $P \propto \Omega^2 \propto \Delta$.

\section{Changes to the quantum motional distribution}
\label{changes}

\begin{figure}
	\input{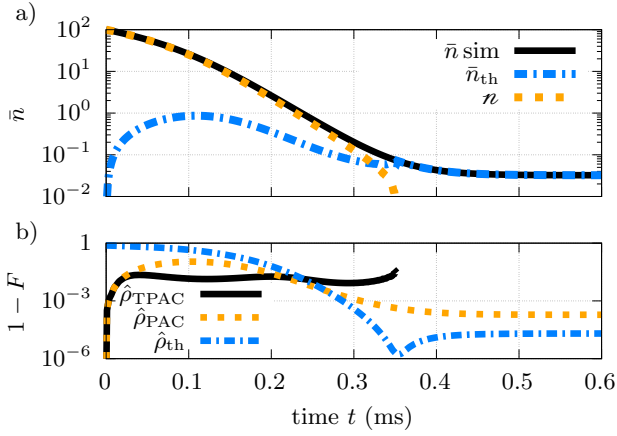}
	\caption{Characterizing the quantum motional distribution $\hat \rho_m$ during the cooling process. a) Decomposing the average occupancy $\bar n$ vs. time $t$ from the fully quantum model (black, solid) with a thermal component $\bar n_{\text{th}}$ (blue, dot-dashed) and a coherent component $\mathscr n$ (orange, dashed). b) Infidelity $1-F$ vs. $t$ [Eq.~\eqref{fidelity}] between the simulated distribution and a TPAC state $\hat \rho_\text{TPAC}$ [Eq.~\eqref{TPAC}, black, solid], PAC state [Eq.~\eqref{PAC}, orange, dashed], and a thermal state $\hat \rho_\text{th}$ [blue, dot-dashed]. Simulation parameters are the same as Fig.~\ref{fig:ratesvsn}.}
	\label{fig:componentsvstime}
 \end{figure}

Our results in Fig.~\eqref{fig:ratesvsn} imply that, for this specific example, the cooling dynamics are parametrized by a single variable: the average occupancy of a coherent state $\mathscr n$.
This parametrization is not surprising early in the cooling process because $\hat \rho_m$ was initialized in a PAC state, $\hat \rho_\text{PAC}(\mathscr n_0 = 100)$.
However, it is not obvious if $\hat \rho_m$ remains in a PAC state $\hat \rho_\text{PAC}[\mathscr n(t)]$ as it is cooled near to the quantum ground state.
Here we determine if the form of $\hat \rho_m$ changes throughout the cooling process, and hence how long the predicted semiclassical cooling rate accurately reflects the true cooling dynamics.

To characterize how $\hat \rho_m$ changes over time, we parameterize it in terms of a more detailed distribution, the thermal phase-averaged coherent (TPAC) state~\cite{tdc}:
\begin{equation}
\label{TPAC}
	\hat \rho_\text{TPAC}(\mathscr n, \bar n_\text{th}) \equiv \frac{1}{2\pi} \int_0^{2\pi} \hat D(\alpha) \frac{e^{- \beta \omega \hat a^\dag \hat a}}{Z} \hat D^\dag(\alpha) \, d \theta.
\end{equation}
Here, $\alpha = \sqrt{\mathscr n} e^{i \theta}$, $\beta \omega = \log[(1+ \bar n_{\text{th}})/\bar n_{\text{th}}]$, and $Z=(1-e^{-\beta \omega})^{-1}$.
This choice is motivated by the fact that $\hat \rho_m$ will diffuse in the neighborhood of the initial PAC state (described by the coherent component $\mathscr n$) due to spontaneous emission and sideband asymmetries, resulting in an additional thermal component $\bar n_{\text{th}}$.
Under the assumption that $\hat \rho_m(t) \approx \hat  \rho_\text{TPAC}(t)$, these components are related to the mean $\bar n$ and variance $\sigma^2$ of $\hat \rho_m$ through the relations
\begin{align}
\label{components}
\begin{split}
	\mathscr n  & \approx \sqrt{\bar n (\bar n + 1) - \sigma^2}; \\
	\bar n_{\text{th}} & \approx \bar n - \mathscr n
\end{split}
\end{align}
at each time $t$.
To further validate this decomposition, we also calculate the fidelities
\begin{equation}
\label{fidelity}
	F(\hat \rho_m, \hat \rho';t) = \left(\text{Tr}\sqrt{\sqrt{\hat \rho_m} \hat \rho' \sqrt{ \hat \rho_m}}\right)^2
\end{equation}
for $\hat \rho' \in \{\hat \rho_\text{TPAC}, \hat \rho_\text{PAC},\hat \rho_\text{th}\}$ at each time $t$.

As shown in Fig.~\ref{fig:componentsvstime}(a), we observe a relatively small, transient thermal component (blue, dot-dashed) as large as $\bar n_{\text{th}} = 1$ at intermediate cooling times, but it remains much smaller than the coherent component (orange, dashed) until the simulated average occupancy (black, solid) has been cooled to $\bar n \ll 1$.
At $t \approx 0.35$ ms, $\mathscr n$ (unphysically) becomes imaginary according to Eq.~\eqref{components}; the simulated distribution is then found to better match a thermal state (blue, dot-dashed), as shown in Fig.~\ref{fig:componentsvstime}(b).
Nevertheless, the TPAC state (black, solid) maintains the highest fidelity when $2 \leq \bar n \leq 100$, indicating that the TPAC state is the most accurate representation of the true quantum motional distribution $\hat \rho_m$ for most of the cooling process.
We conclude that $\hat \rho_m$ remains approximately in a TPAC state with $\bar n_\text{th} \ll \mathscr n$ until it approaches steady-state, confirming that the distribution is parameterized by $\bar n(t) \approx \mathscr n(t)$ and that the semiclassical cooling rates in Fig.~\ref{fig:ratesvsn} are an accurate predictor of the dynamics over the cooling process.

We emphasize that our investigation here is not exhaustive, and there are cases such that the form of the quantum motional distribution is substantially altered by the cooling, thereby limiting the amount of time we can accurately extrapolate the semiclassical prediction.
For example, the experimental results in Section~\ref{expComparison} suggest that the modes thermalize before reaching steady state; however, the mode was initialized in a much higher energy coherent state $\mathscr n_0 \approx 1000$, and the richer internal structure may have resulted in some additional heating.
It may then be necessary to resort to other approaches, such as a Fokker-Planck formalism.
Nevertheless, our approach offers an efficient model for predicting cooling rates in all energy regimes at early cooling times.

\section{Derivation of cooling dynamics for phase-averaged quantum motional distributions}
\label{instantaneousRatesAppendix}

We show here how to use PACMAN to predict cooling dynamics for a phase-averaged motional distribution $\hat \rho$.
More specifically, we calculate the time derivatives for each moment $\braket{\hat n_\mu^k}$ from the semiclassical cooling rates, where $k = 1,2, \ldots$ is the moment order and $\mu = 1,2,...,N$ is the mode index.

Because the semiclassical approach yields cooling predictions for coherent states, it is natural to express $\hat \rho$ as
\begin{equation}
\label{Prep}
	\hat \rho = \int d^{2N} \bs \alpha \ket{\alpha} \bra{\alpha}^{\otimes N} P(\bs \alpha)
\end{equation}
in which $P$ is the distribution used in the Glauber–Sudarshan P representation~\cite{carmichaelvol1}.
Here we have used the notation $\bs \alpha = \alpha_1, \ldots, \alpha_N$ and 
\begin{equation}
	 d^{2N} \bs \alpha = d^2 \alpha_1 \ldots d^2 \alpha_N.
\end{equation}
Because $\hat \rho$ is phase-averaged, then $P$ satisfies $P(\bs \alpha) = P(|\bs \alpha|)$, in which $|\bs \alpha| = |\alpha_1|, |\alpha_2|,\ldots$. 
By parameterizing each coherent state eigenvalue as $\alpha_\mu = |\alpha_\mu| e^{i \theta_\mu}$, we carry out the angular integrals to obtain the simpler expression
\begin{equation}
\label{Prep_symmAppendix}
	\hat \rho = \pi^N \int d^N \bs{\mathscr n} \, \hat \rho_\text{PAC}^{\otimes N}(\bs{\mathscr n}) \, P (\bs{\mathscr n}).
\end{equation}
Here $\mathscr n_\mu \equiv |\alpha_\mu|^2$ is the average occupancy of mode $\mu$ in a coherent state, and
\begin{equation}
 \hat \rho_\text{PAC}^{\otimes N}(\bs{\mathscr n}) = \prod_\mu \hat \rho_\text{PAC}(\bs{\mathscr n}_\mu)
\end{equation}
is the $N$-mode PAC state [see Eq.~\eqref{PAC}].
Using the operator expectation value equation
\begin{equation}
	\braket{ \hat O} = \pi^N \int d^N \bs{\mathscr n} \, \tr{\hat \rho_\text{PAC}^{\otimes N}(\bs{\mathscr n}) \hat O} \, P(\bs{\mathscr n}),
\end{equation}
we calculate the single-mode moments to be
\begin{equation}
\label{moments}
	\braket{\hat n_\mu^k} = \pi^N \int d^N \bs{\mathscr n} \, \mathscr n_\mu^k \, P(\bs{\mathscr n}).
\end{equation}

It can be shown that $P(\bs{\mathscr n})$ evolves according to a Fokker-Planck equation~\cite{carmichaelvol1,highTempOscillator}:
\begin{equation}
\label{FP}
	\frac{\partial P}{\partial t} = 
		\sum_\mu \frac{\partial}{\partial \mathscr n_\mu} \left[ \mathscr n_\mu \, W_{\text{SC}, \mu} (\bs{\mathscr n}) P(\bs{\mathscr n})\right].
\end{equation}
Here we have neglected diffusive heating, applied the rotating wave approximation to remove terms oscillating at the (relatively fast) mode frequencies $\omega_\mu$, and inserted the semiclassical cooling rates $W_{\text{SC}, \mu}(\bs{\mathscr n})$ (as opposed to one calculated through a more quantum treatment~\cite{rabl2010,highTempOscillator}).
The evolution of the entire distribution can be determined at all times by solving the Fokker-Planck equation, but this may be computationally expensive.
Differentiating Eq.~\eqref{moments} in time, substituting in~\eqref{FP}, and applying integration by parts, we find
\begin{equation}
\label{dndtPACppendix}
	\frac{d}{dt} \braket{\hat n_\mu^k}
		= - k \pi^N \int d^N\bs{\mathscr{n}} \,  \mathscr n_\mu^k \, W_{\text{SC}, \mu}(\bs{\mathscr n}) \, P(\bs{\mathscr n}).
\end{equation}
We have dropped the boundary terms with $\mathscr n_\mu \rightarrow \infty$ and all other terms for modes $\nu \neq \mu$ since $ P(\bs{\mathscr n})$ is normalized.
This formalism allows us to use $W_{\text{SC}, \mu}(\bs{\mathscr n})$ to predict changes to any phase-averaged quantum motional distribution due to laser cooling.

\end{document}